\documentclass[fleqn,usenatbib]{mnras}

\usepackage{newtxtext,newtxmath}
\usepackage[T1]{fontenc}
\usepackage{ae,aecompl}

\usepackage{graphicx}	% Including Figure~files
\usepackage{amsmath}	% Advanced maths commands
\usepackage[shortlabels]{enumitem}
\usepackage{booktabs}

\usepackage{multirow}
\usepackage[table,xcdraw]{xcolor}
\usepackage{tabularx}

\defcitealias{salt3}{K21}
\defcitealias{jones_22_host}{J23}
\defcitealias{bs20}{BS21}

\title[Custom Light Curve Models of Type Ia Supernova Sub-Populations]{Training Custom Light Curve Models of SN~Ia Sub-Populations Selected According to Host Galaxy Properties}

\author[G. Taylor]{
G. Taylor,$^{1}$\thanks{E-mail: georgina.taylor@anu.edu.au}, 
C. Lidman,$^{1}$ 
B. Popovic,$^{2}$
H. J. Abbot$^{1}$ 
\\
$^{1}$Research School of Astronomy and Astrophysics, Australian National University, Canberra, Australia\\
$^{2}$Univ Lyon, Univ Claude Bernard Lyon 1, CNRS, IP2I Lyon/IN2P3, IMR 5822, 69622 Villeurbanne, France}

\date{Accepted XXX. Received YYY; in original form ZZZ}

\pubyear{2023}

\begin{document}

\label{firstpage}
\pagerange{\pageref{firstpage}--\pageref{lastpage}}
\maketitle

%It should be a single paragraph not more than 250 words (200 words for Letters).
%No references should appear in the abstract.
\begin{abstract}
Type Ia supernova (SN~Ia) cosmology analyses include a luminosity step function in their distance standardization process to account for an observed yet unexplained difference in the post-standardization luminosities of SNe~Ia originating from different host galaxy populations (e.g., high-mass ($M \gtrsim 10^{10} M_{\odot}$) versus low-mass galaxies). We present a novel method for including host-mass correlations in the SALT3 light curve model used for standardising SN Ia distances. We split the SALT3 training sample according to host-mass, training independent models for the low- and high-host-mass samples. Our models indicate that there are different average Si II spectral feature strengths between the two populations, and that the average SED of SNe from low-mass galaxies is bluer than the high-mass counterpart. We then use our trained models to perform a SN cosmology analysis on the 3-year spectroscopically confirmed Dark Energy Survey SN sample, treating SNe from low- and high-mass host galaxies as separate populations throughout. We find that our mass-split models reduce the Hubble residual scatter in the sample, albeit at a low statistical significance. We do find a reduction in the mass-correlated luminosity step but conclude that this arises from the model-dependent re-definition of the fiducial SN absolute magnitude rather than the models themselves. Our results stress the importance of adopting a standard definition of the SN parameters ($x_0, x_1, c$) in order to extract the most value out of the light curve modelling tools that are currently available and to correctly interpret results that are fit with different models.

\end{abstract}


%%%%%%%%%%%%%%%%%%%%%%%%%%%%%%%%%%%%%%%%%%%%%%%%%%

%%%%%%%%%%%%%%%%% BODY OF PAPER %%%%%%%%%%%%%%%%%%

\section{Introduction}
Type Ia supernovae (SNe Ia) are standardisable candles whose distances can be measured to within $\sim5\%$ via the classical two-parameter standardisation process \citep{tripp98}.\footnote{Some studies have suggested that SN observations made in the near-infrared are consistent enough to yield distance measurements without the need for standardisation parameters \citep{mandel_09, johansson_21}; however, the scarcity of near-infrared data means that this technique has not yet been widely adopted.} These distance measurements can be made out to $z~\sim~2.2$ with current observing facilities \citep{Jones_2013_highz, Graur_2014_highzSN, Rodney_2014_highzSN, Riess_2018_highzSN, hayden_2021}, making SNe Ia a uniquely useful class of object for studying the history of the expansion of the Universe and phenomena such as dark energy 
\citep{g10_scatter, suzuki12, jla, dessnconstraints19, panstarrs18, Jones_2019, panplus_constraints}. Low redshift SNe~Ia are also useful for making local measurements of the Hubble constant \citep{branch_98_h0,riesshubb, galbany_22, riess_22}. 

The SN parameters necessary for standardising distances are obtained by fitting a light curve model to observed time-series photometry of a sample of SNe~Ia. For the last decade, SALT2 has been the most widely used light curve model \citep{salt2, g10_scatter, mosher14, jla, T21, superfrag, t23}. Recently, \citet{salt3} introduced SALT3, a modern refactored version of the SALT2 model. SALT3 has a number of advantages over SALT2 \citep{salt3, salt3nir, t23}. In particular, SALT3's actively maintained, well-documented, open-source training procedure makes it straightforward for the SN community to use and update SALT3 as we improve training data sets and gain a deeper understanding about the relationship between the model inputs and cosmological constraints \citep[e.g., ][]{jones_22_host, dai_22}.

An open question in the measurement of current SN distances (and moreover, cosmological constraints) is the true nature of the ``mass step". The mass step is an observed $\sim 0.06$ magnitude shift in standardised SN~Ia distance moduli as a function of host galaxy mass \citep{kelly2010, lampeitl_2010, sullivan_host, smith2020cosmology, kelsey_2021}. It is observed to be more significant for redder SNe~Ia \citep{bs20, kelsey_23}. The astrophysical origins of this effect are as yet unconfirmed; the step in luminosity has also been observed to correlate with other parameters, such as progenitor metallicity \citep{D_Andrea_2011, hayden_13, rose_21_mass}, progenitor ages (\citealt{childress_14}), specific star formation rate (i.e., the star formation rate per unit of stellar mass, be it local: \citealt{rigault_2013, 393, kim_18, rigault_20, briday_22, hand_22}; or global: \citealt{D_Andrea_2011, hand_22}), age of the stellar population \citep{Rose_2019}, host galaxy morphology \citep{Pruzhinskaya_20}, local host colour or mass \citep{roman_18, jones-18_local, kelsey_2021}, local dark matter density \citep{steigerwald_22}, or host galaxy dust \citep{bs20, Popovic_2021, wiseman_22}. However, it is still most common for SN cosmology analyses to standardise their distances with a mass step \citep{conley11, jla, panstarrs18, dessnconstraints19, Jones_2019, panplus_constraints} --- although \citep{bs20} suggest that the observed step in SN luminosities is actually caused by differences in galactic dust laws that happen to correlate with host galaxy mass, and that by accounting for the diversity of dust properties, the need for the mass-step disappears.

Currently, the mass step is accounted for in SN cosmology analyses by deriving distances from a model (such as SALT2 or SALT3) that does \textit{not} parameterise any host galaxy relationship, and then measuring how the Hubble residuals correlate with host properties in order to make a luminosity correction. For example, in the Dark Energy Survey 3-Year Supernova Analysis (DES-SN3YR; \citealt{dessnconstraints19}) the host-mass correction is added to the standard Tripp equation \citep{tripp98} to measure distance moduli:
\begin{equation}
     \label{eq:modified_tripp_p3}
 	\mu = m_b - M +\alpha x_1 - \beta c + \delta \mu_{\text {host }} + \delta_{\mu_{\rm bias}} ,
\end{equation}
where $\mu$ is the distance modulus of a SN~Ia in magnitude units; $M$ is the absolute magnitude of a fiducial ($x_0 = 1, x_1=0, c=0$) SN~Ia (and is degenerate with the Hubble constant $H_0$);\footnote{Following \citep{salt2mu}, $M$ is offset from the absolute $B$-band magnitude by $\sim 10$ mag.} $\alpha$ and $\beta$ are global dimensionless nuisance parameters representing the amplitude of the stretch-luminosity and colour-luminosity relationships; $m_b$, $x_1$ and $c$ are respectively the apparent magnitude at peak, stretch and colour parameters of an individual SN (recovered from the SALT3 light curve fit); and $\delta_{\mu_{\rm bias}}$ accounts for bias corrections. The host-mass correction is:
\begin{equation}
\label{mass_step_eq}
\delta \mu_{\mathrm{host}}=\gamma \times\left[1+e^{\left(M_{\log \mathrm{host}}- M_{\log \mathrm{step}}\right) / \tau_{M_\star}}\right]^{-1}-\frac{\gamma}{2},
\end{equation}
where $M_{\log \mathrm{host}} = \log_{10}{M_\star}/M_{\odot}$ represents the stellar mass of the host galaxy, $M_{\log \mathrm{step}}$ is the chosen ``split" between high-mass and low-mass galaxies (typically set to be $10$), $\tau_{M_\star}$ is the width of the mass split, and $\gamma$ is the mass step which is fit (e.g., \citealt{des3yr_systematics}). 

Regardless of its origins, it is now possible to explore the cosmological impact of the mass step in the underlying light curve model. By leveraging its training code accessibility and the large SNe~Ia training samples that are now available, SALT3 can offer a insight into the mass step by revealing differences in the spectral energy distributions (SEDs) between high-host-mass and low-host-mass populations of SNe~Ia. A SALT3-based investigation into the impact of the mass step was first made in \citet{jones_22_host} (hereafter \citetalias{jones_22_host}). They added a mass step parameterization to the model, finding differences in some spectral features; most notably in the equivalent widths of Ca H\&K and Si II (at $\sim 2\sigma$ significance). While using the \citetalias{jones_22_host} model parameterisation for light curve fitting does reduce the mass step by $0.021\pm0.002$ mag, they find little change to the Hubble residual dispersion. In order to ensure robust host galaxy masses, \citetalias{jones_22_host} only used 278 low-$z$ ($z < 0.15$) SNe~Ia to train its models --- just 30\% of the SALT3 training sample used in \citealt{salt3} (hereafter \citetalias{salt3}).

Here, we build on the investigation of \citetalias{jones_22_host} by splitting the complete \citetalias{salt3} SALT3 training sample of 1083 SNe~Ia on host galaxy mass. We take a different approach to \citetalias{jones_22_host}, choosing to use the traditional two-parameter SALT3 model (without any host galaxy considerations) and simply train separate SALT3 surfaces (i.e., model iterations) from a ``high-host-mass" ($M_{\rm host} > 10^{10} M_\odot$) training sample and ``low-host-mass" sample. In this way, we can see how all SALT3 components change with host-mass, rather than parameterising the impact of host-mass within a single SED component. We focus on  comparing the results on a cosmological level by performing a reanalysis of the spectroscopically confirmed Dark Energy Survey 3 Year SN data (\citealt{deslc, des3yr_systematics, dessnconstraints19}).

The SALT3 model training procedure underlying this work is outlined in $\S$\ref{s3_training}. Our methods for training SALT3 models on samples with different host-mass distributions are given in $\S$\ref{sec:mass_step}. We present and discuss all results in $\S$\ref{sec:results}. Our conclusions are presented in $\S$\ref{sec:summary}.

\section{SALT3 Model Training}
\label{s3_training}

The SALT3 model spectral flux density at phase $p$ and wavelength $\lambda$ is defined as:
\begin{align}
\label{eq:salt_flux_p3}
    F_{\lambda}= x_{0} &\times\left[M_{0}(p, \lambda)+x_{1} M_{1}(p, \lambda)\right] \times \exp [c C L(\lambda)],
\end{align}
where $x_{0}, x_{1}$ and $c$ are the parameters of a particular supernova and $M_0$, $M_1$, and $C_L$ are the global model components. These components are iteratively fit using a $\chi^2$ minimization process during the model training.

The SALT3 training process is outlined in \citetalias{salt3}. This process requires an input training sample of photometric and spectroscopic data from well-measured SNe~Ia, along with information about the systems on which the photometry was observed, the redshifts, and a set of starting guesses for the SALT3 model components and training sample light curve parameters. We use the the public \verb|SALTShaker| training code with the default public \citetalias{salt3} training configuration in this work, and select our training subsamples from the \citetalias{salt3} training sample.\footnote{During the preparation of this manuscript, a new version of SALTShaker that utilises JAX gradient descent optimisation was released (\url{https://github.com/djones1040/SALTShaker/releases/tag/v1.1}. This version is faster and has fewer convergence issues, but has not been publicly validated yet \citep{kenworthy_23}.} Following \citet{t23}, we use the photometric calibration presented in \citet{superfrag}.\footnote{Following \citet{t23}, some of the photometry is also updated from the original \citetalias{salt3} release.} We make the following modifications to the default SALT3 configuration:
\begin{itemize}
    \item The number of component fitting iterations between each error fitting iteration is changed from 5 to 10, for computational speed. The error estimation is the most computationally-intensive part of the training process, and this choice halves the time spent in that training stage. This is not anticipated to yield any noticeable difference to the model surfaces \citep{saltshaker_doc}.
    \item Following \citet{dai_22}, we reduce the maximum number of spectral calibration parameters from 10 to 5. During the training, spectra are iteratively recalibrated to match the best-fit model. This accounts for any poor-quality calibration in the spectral training sample, which often has large uncertainties. However, using too high an order for the recalibration polynomial may fit out real spectral features.
    \item The spectral $\chi^2$ scaling, which ensures roughly equal photometric and spectroscopic contributions to the final model reduced $\chi^2$, is fine tuned for each surface such that the contributions are within 10\%. This procedure, which follows \citet{salt2, salt3}, ensures that the spectroscopic and photometric training data are considered with equal weighting, despite the spectroscopic uncertainties often being much larger.
    \item SALTShaker defines the training convergence to occur when the change in $\chi^2$ between consecutive iterations is less than 1. We find that for some of the models trained in this work, this criteria is never reached. In this case, we consider a model to have ``converged" if the change in $\chi^2$ is less than 0.01\%.
\end{itemize}

\section{Investigating Host Galaxy Mass in the SALT3 Model}
\label{sec:mass_step}

Here, we investigate the possibility that separate intrinsic populations of SNe~Ia exist, and are correlated with host galaxy mass. We perform separate light curve modelling for SNe~Ia in ``low-mass" and ``high-mass" host galaxies. We split our populations using host stellar mass, as this is the most commonly used proxy for the host galaxy-luminosity relationship in SN cosmology analyses --- SNe in high-mass hosts are observed to have brighter post-standardization luminosities than those in low-mass hosts.\footnote{This trend is reversed before standardization, where SNe in high-mass hosts (with less active star formation) tend to be dimmer \citep{childress_14, hand_22}.} The methods developed in this work could be applied to other host galaxy parameters of interest (e.g., host galaxy colour, age, morphology, specific star formation rate), provided that there exists sufficient data to train the models.

\subsection{SALT3 Training Samples and Resultant Surfaces}
\label{masssplit_tsamples}

\begin{figure*}
  \begin{centering}
   \includegraphics[width=1.9\columnwidth]{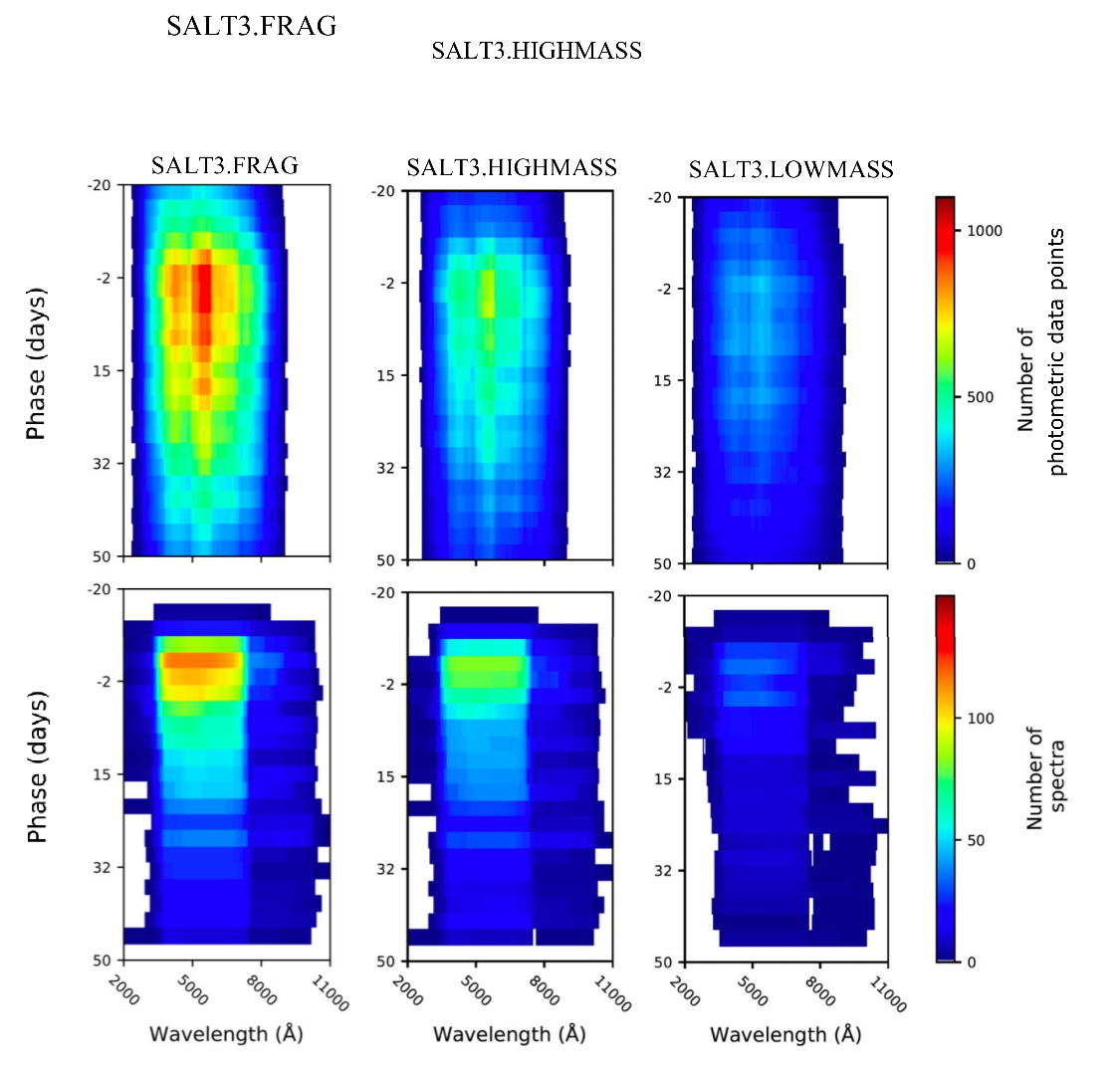}
  \caption{Data density plot showing the phase-wavelength coverage of the full \citetalias{salt3}, SALT3.HIGHMASS, and SALT3.LOWMASS training samples, applying a consistent heatmap scaling. The upper panels show the coverage of the photometric data, and the lower panel shows the spectral coverage. The relative distribution of data between samples is broadly similar, with slightly higher spectroscopic coverage in SALT3.HIGHMASS (which has a higher proportion of low-redshift SNe). These plots were generated using \texttt{SALTShaker}.}
  \label{full_dd_p3}
  \end{centering}
\end{figure*}

\begin{table}
\begin{minipage}{\columnwidth}
  \renewcommand{\arraystretch}{1.5}
  \centering
  \caption{Host galaxy mass statistics for the mass-split SALT3 training samples used in this work. The low-redshift subsample tends to include more high-mass host galaxy SNe, while the untargetted high-redshift subsample favours low-mass hosts.}
    \label{mass_sample_info}
\begin{tabular}{@{}llrrr@{}}
\toprule
\textbf{Sample} & \textbf{Sub-sample} & \textbf{\# SNe} & \textbf{mean $M_{\log}$} & \textbf{median $M_{\log}$}\\ \midrule
\citetalias{salt3} & Full & 1083 & 9.843 & 10.227\\
\citetalias{salt3} & $z < 0.1$ & 337 & 10.075 & 10.452\\
\citetalias{salt3} & $z > 0.1$ & 748 & 9.729 & 10.081\\
TS-HIGHMASS & Full & 617 & 10.709 & 10.667\\
%TS-HIGHMASS & $z < 0.1$ & 227 & 10.672 & 10.650\\
%TS-HIGHMASS & $z > 0.1$ & 390 & 10.730 & 10.702 \\
TS-LOWMASS & Full & 446 & 8.645 & 9.189\\
%TS-LOWMASS & $z < 0.1$ & 92 & 8.717 & 9.130\\
%TS-LOWMASS & $z > 0.1$ & 354 & 8.626 & 9.214\\
\bottomrule
\end{tabular}
\end{minipage}
\end{table}

\begin{figure*}
  \begin{center}
   \includegraphics[width=0.7\textwidth]{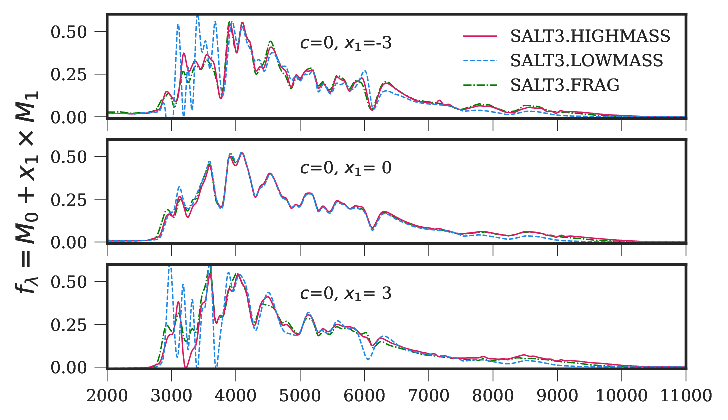}
  \end{center}
  \caption{The combined $M_0$ and $M_1$ components of SALT3 at various $x_1$ values for the trained SALT3.LOWMASS and SALT3.HIGHMASS surfaces, as well SALT3.FRAG \citep{t23}. There is clear evidence of ringing (rapid fluctuations in phase or wavelength space) in the UV region of the SALT3.LOWMASS $M_1$ component, which limits its usefulness in this analysis.}
  \label{sed_ringing}
\end{figure*}

\begin{figure*}
  \begin{center}
   \includegraphics[width=0.75\textwidth]{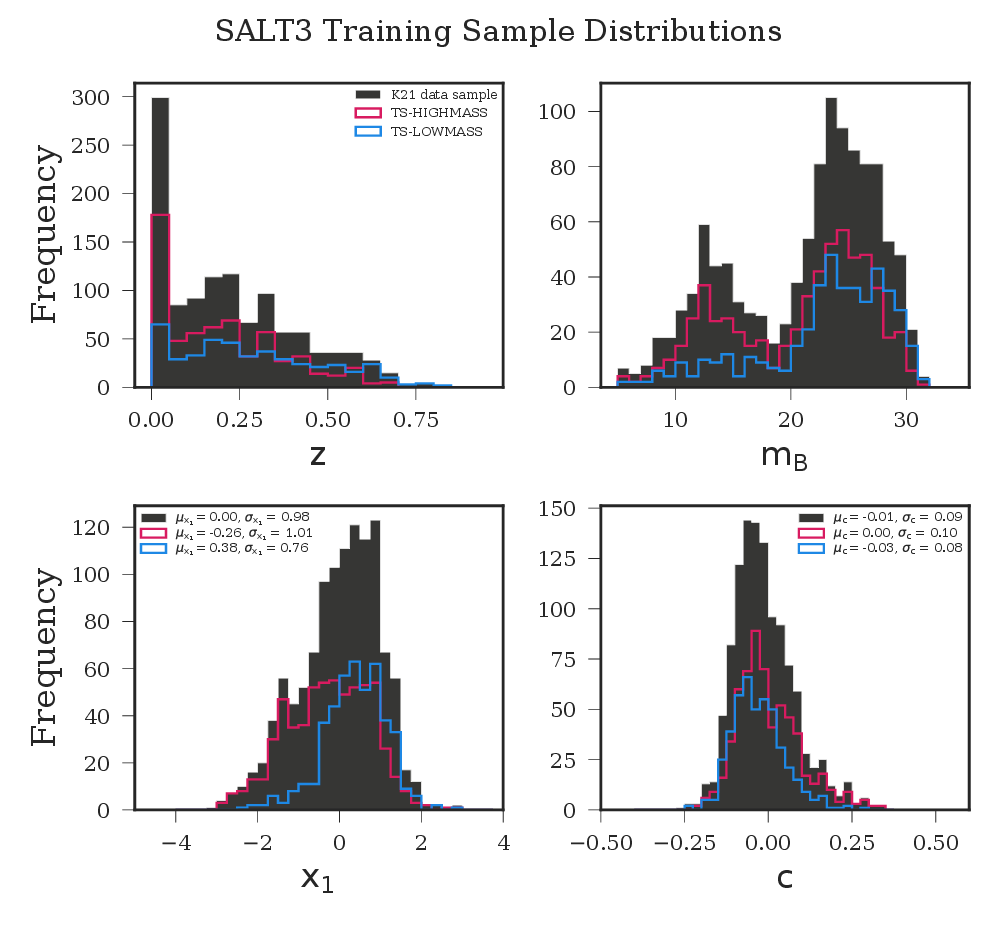}
  \end{center}
  \caption{SN parameter distributions of the mass split training samples used in this work. Here, we plot $m_B = -2.5\ln(x_0)$ as a measure of business. The parameters are all obtained from light curve fits with the same SALT3 model, and are used as initial guesses in the \texttt{SALTShaker} training procedure. While the colour distributions are similar between the high- and low-host-mass training samples, the amplitude and stretch distributions change significantly.}
  % from Assemble_Host_Mass_Table notebook (14 Mar 23)
  \label{mass-dist}
\end{figure*}

We use host-masses from the public \citetalias{salt3} data files, which were provided for 662 of the 1083 SNe. The missing masses were primarily from SNe in the Sloan Digital Sky Survey (SDSS, \citealt{holtzman_08}), Supernova Legacy Survey (SNLS, \citealt{snls06}), and various low-$z$ surveys \citep{jha_06}. We supplement the \citetalias{salt3} host-mass information with data from SNLS (M. Sullivan 2021, private communication) and the Joint Light Curve Analysis (JLA, \citealt{jla}). This gives us mass information for all but 20 SNe in the K21 sample. We split the \citetalias{salt3} sample to construct two training sub-samples: \textit{TS-HIGHMASS}, consisting of SNe in ``high-mass" ($M_{\log} > 10$) host galaxies, and \textit{TS-LOWMASS}, consisting of SNe in ``low-mass" ($M_{\log} < 10$) host galaxies. We apply the mass-split at $M_{\log} = 10$ (where $M_{\log} = \log_{10} M_\star /M_\odot$) for consistency with the literature; this is slightly higher than the mean $M_{\log} = 9.483$ of the full sample. Summary statistics of the training samples are given in Table~\ref{mass_sample_info}.

Unlike \citetalias{jones_22_host}, we have not derived our own host galaxy masses. It's therefore possible that using multiple sources of host galaxy mass could introduce some errors. However, as we are only applying a broad cut at $M_{\log}$ = 10 and do not use the host galaxy mass information in the actual training process, our masses need only be accurate enough to assign each SN to the appropriate high- or low-host-mass bin. The \citetalias{salt3} training sample has a mean low-host-mass error of $\sigma^{\rm low}_{M_{\rm log}} = 0.661$ and a mean high-host-mass error of $\sigma^{\rm high}_{M_{\rm log}} = 0.469$.\footnote{\citetalias{jones_22_host} found that the uncertainty on their host galaxy masses was small enough to be negligible, and so did not propagate that uncertainty into their SALT3 model uncertainty.}

Making use of publicly available host-masses allows us to use nearly all of the \citetalias{salt3} training sample SNe, while \citetalias{jones_22_host} could only use a fraction of the low-$z$ SNe. It is important to include as many SNe in the SALT3 training as one reliably can. The training of SALT models will only converge if there is adequate phase-wavelength coverage in the training sample. Even if the training does converge, low-coverage regions of the phase-wavelength space may be poorly constrained or suffer from training artefacts that require tailored regularization. We do not adjust the regularization parameters in this work, instead choosing to only focus on results obtained from well-constrained regions of the trained models.

In order to assess the reliable regions of our trained models, we provide phase-wavelength coverage plots for each of the training samples (Figure~\ref{full_dd_p3}). TS-HIGHMASS largely follows the phase-wavelength space footprint of the \citetalias{salt3} sample, with only minor gaps in some late-phase spectral coverage. The density of the data coverage is at least as good as that of the Joint Light curve Analysis training sample from \citet{jla} (shown in Figure~10 of \citetalias{salt3}) which was used to produce the SALT2.4 ``JLA" model --- the most widely used model in the literature \citep[e.g.,][]{jla, dessnconstraints19, panstarrs18}. We are therefore confident that our high-host-mass training sample is sufficient to produce a reliable SALT3 model (``SALT3.HIGHMASS").

While the relative phase and wavelength coverage of TS-lowmass broadly matches that of the \citetalias{salt3} sample, the number of SNe is lower. It is even lower than the JLA sample, with the maximum density bins reduced by $\sim$50\%. This follows directly from the $\sim$50\% reduction in the number of SNe in TS-LOWMASS. The spectral coverage is also slightly poorer: there are gaps at multiple phases in the rest-frame ultraviolet (UV, $\lambda \lesssim4000$\AA) and near-infrared (NIR, $\lambda \gtrsim9000$\AA) regions --- though the spectroscopic coverage matches JLA reasonably well for the other regions.\footnote{We note that the original JLA sample also did not include spectra at NIR wavelengths, which are not typically used for light curve fitting anyway.} Given this limited data, we carefully inspect the resultant surface (``SALT3.LOWMASS") components for ringing artefacts. We find evidence of significantly ringing in the $M_1$ component (and mild ringing in $M_0$) at many phases, including peak, for $\lambda \lesssim4000$\AA\ (e.g., Figure~\ref{sed_ringing}). At longer wavelengths, our SALT3.LOWMASS surface appears to be smoothly constrained (there is a peculiar $M_1$ feature around 6000\AA, but we find no evidence to suggest this is a training artefact). Ameliorating the UV ringing artefact would require a refined SALT3 regularization scheme or a supplementary training sample of SNe in low-mass host galaxies, which are beyond the scope of this paper. Instead, we quantify the impact of the UV ringing by performing two identical cosmological analyses over different rest-frame wavelength ranges, that either include or exclude the affected region of the SALT3.LOWMASS SED.

A further limitation in our analysis relates to scaling employed by SALT3.\footnote{SALT2 also employs a scaling of SN parameters in its training process.} The SALT3 training performs a scaling of the SN parameters, such that the training sample has $\bar{x_1} = 0,\ \sigma_{x_1} = 1,\ \bar{c} = 0$ (\citetalias{salt3}, \citealt{dai_22}). Similarly, the SALT3 components are defined relative to the training sample demographics: $M_0$ is the average SED for a $x_1 = 0,\ c = 0$ SN, and $M_1$ scales with $\sigma_{x_1}$. Assuming our input training sample distributions are representative of the \textit{true} $x_1$ and $c$ populations of SNe in high- and low-mass host galaxies as they occur in the Universe, any differences in the $M_0$ components should represent the average astrophysical differences between SNe originating from high- and low-mass host galaxies. However, any biases in the input training sample distribution will complicate this interpretation. The end result is that the fitted light curve parameters from each model cannot be meaningfully compared without a scaling correction. We therefore compare the SN parameters that are determined from fitting a single SALT3 surface to the entire training sample. We plot these parameter distributions for each of our mass-split training samples in Figure~\ref{mass-dist}. While the $c$ distributions are largely consistent across our training samples, there is (unsurprisingly) a significant change in the $x_1$ distributions --- SNe~Ia that occur in more massive galaxies tend to be faster declining, so have lower $x_1$ values \citep{hamuy_95, branch_96, hamuy00, howell_01, gallagher_2005}.\footnote{A relationship with colour has also been found, though as we see in Figure~\ref{mass-dist}, this are weaker than the stretch relationships \citep{smith2020cosmology}.} Moreover, the low-mass training sample has a narrower range of $x_1$ values than the high-mass or full training samples.

\subsection{Methods to Assess the Impact on Distances and Cosmology}
\label{p3a_methods}

We use the Dark Energy Survey's 3-year spectroscopically confirmed SNe~Ia sample and its companion low-$z$ sample \citep[DES-SN3YR][]{deslc} to test the end-to-end cosmological impact of treating SNe~Ia from different host populations as intrinsically different. We fit the DES-SN3YR light curves with our SALT3.HIGHMASS and SALT3.LOWMASS surfaces, as well as with a control surface, SALT3.FRAG \citep{t23}, which was trained with the same inputs but uses the entire \citetalias{salt3} training sample. We split the DES-SN3YR sample SNe according to the surface demographics --- for example, we fit the entire sample with SALT3.FRAG, but only the $M_{\log} > 10 M_{\odot}$ subsample with SALT3.HIGHMASS. We use the published SVA1-Gold DES-SN3YR host-masses to select these subsamples \citep{dessnconstraints19}, as these were the masses used in the \citetalias{salt3} training sample. The high-host-mass subsample contains 91 DES SNe and 98 low-$z$ SNe; the low-host-mass subsample contains 116 DES SNe and 24 low-$z$ SNe. There are 18 SNe in the DES sub-sample for which hosts could not be adequately identified. Following \citet{des3yr_systematics}, these SNe are assumed to be in low-mass hosts (as high-mass hosts would likely be detectable). There are 10 ``high-host-mass" DES SNe within $1\sigma_{M_{\log}}$ of our chosen mass boundary ($M_{\log} = 10$), and 6 ``low-host-mass" DES SNe within $1\sigma_{M_{\log}}$ of the boundary. There are 28 high-host-mass, low-$z$ SNe within $1\sigma_{M_{\log}}$ of our chosen mass boundary, and 11 low-host-mass, low-$z$ SNe within $1\sigma_{M_{\log}}$ of the boundary.

We use \texttt{SNANA} \citep{snana09} to fit the light curves over $3000 < \lambda_{\rm eff} < 8000$\AA, where $\lambda_{\rm eff}$ is the rest-frame effective mean wavelength of the photometric filter. In this range, $U$-band observations in low-$z$ SNe (and $g$ band observations at $z \gtrsim 0.3$) will be impacted by the UV ringing artefact seen in the SALT3.LOWMASS surface (Figure~\ref{sed_ringing}). We explore the impact of this ringing by performing a similar analysis over a truncated rest-frame wavelength range ($4000 < \lambda_{\rm eff} < 8000$\AA); this is presented in Appendix~\ref{4000_analysis}.

 We apply standard light curve cuts of $-0.3 < c < 0.3$, $-3 < x_1 < 3$ following \citet{des3yr_systematics, jones_22_host}. We use the \verb|SALT2mu| procedure of \citet{salt2mu} to fit nuisance parameters and distance moduli for three sets of light curve fit results --- the full DES-SN3YR sample (including low-$z$) fit with SALT3.FRAG, the high host-mass SNe fit with SALT3.HIGHMASS, and the low host-mass SNe fit with SALT3.LOWMASS. For these three sets of results, we include 1D bias corrections but do not include any host-mass correction ($\delta \mu_{\text {host }}$, Equation~\ref{eq:modified_tripp_p3}). We fit an additional set of distance moduli (labelled SALT3.FRAG$ + $Mass Step) that uses light curve parameters from the full DES-SN3YR sample fit with SALT3.FRAG and \textit{does} include the fit of a mass step term, with $M_{\log \mathrm{step}} = 10$. 
These results use the same set of bias correction simulations as the nominal SALT3.FRAG results, and yield a $\gamma$ value (Equation~\ref{mass_step_eq}) that is comparable to the 1D mass step found in \citep{smith2020cosmology} ($\gamma_{\rm{Smith2020}}$ = 0.066 $\pm$ 0.020mag, $\gamma_{\rm{Taylor2023}}$ = 0.079 $\pm$ 0.021mag).

The 1D bias corrections are calculated using large sets of realistic simulations of the DES-SN3YR data, which are generated following \citet{Kessler_2019_sims}. Separate simulations are created for the low-$z$ and DES SNe, for each SALT3 surface. The simulated SN are drawn from parent populations that are calculated according to predicted distributions from each individual surface, following the methods of \citet{Popovic_2021}. While there are no explicit mass-luminosity relationships included in our simulations, each set selects SNe from parent populations that are calculated according to the corresponding real distribution of fitted DES-SN3YR data. For example, the SALT3.HIGHMASS simulations are generated from stretch and colour distributions that are calculated using only the high host-mass DES-SN3YR SNe fit with the SALT3.HIGHMASS model. Each set of bias correction simulations also uses the corresponding SALT3 model as the input model to generate the rest-frame SED. We use the G10 intrinsic scatter model of \citet{Kessler_2013}, based on \citet{g10_scatter}, and assume an underlying flat $\Lambda$CDM cosmology with $\Omega_M = 0.315, \Omega_\Lambda = 0.685, H_0 = 70$km/s/Mpc. These assumptions introduce a small systematic uncertainty in the recovered cosmological parameters (Table 8 of \citealt{des3yr_systematics}), though we only include statistical uncertainties in our results here.

For each set of results, we fit $\Omega_m$ and $w$ assuming a flat $w$CDM model and applying a CMB prior, tuned such that the constraining power is similar to that of Planck \citep{planck2015}. This is performed with \verb|SNANA|'s fast cosmology fitting program, \verb|wfit|. We verify our SALT3.HIGHMASS and SALT3.LOWMASS surfaces' ability to recover accurate cosmological parameters by using them to fit simulated DES-SN3YR light curves that are generated with a known cosmology. In these test, we recover the input $\Omega_m$ and $w$ values to within $2\sigma$ (when applying a CMB prior). 

We also test how much of the difference in results from our three surfaces can be attributed to the surfaces themselves, versus the use of different subsamples of  DES-SN3YR SNe. We do this by fitting the high/low-host-mass DES-SN3YR subsample light curves with SALT3.FRAG, and using each set of results to obtain separate distance measurements and cosmological fits. The results of these tests (given in $\S$~\ref{sec:distance_results}) are labelled ``SALT3.FRAG (HIGH)" and ``SALT3.FRAG (LOW)". There is no explicit mass-step applied in this test, as we expect any luminosity difference between the subsamples to be absorbed into the $M$ parameter of Equation~\ref{eq:modified_tripp_p3}. 

\section{Results and Discussion}
\label{sec:results}
\begin{figure*}
  \begin{center}
   \includegraphics[width=0.75\textwidth]{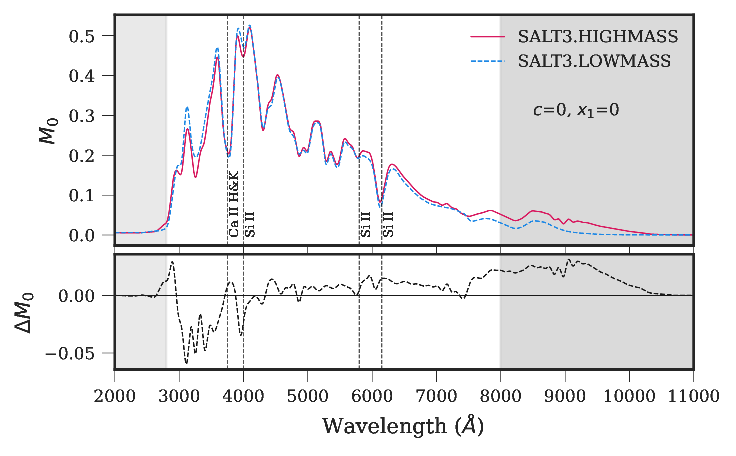}
  \end{center}
  \caption{The $M_0$ components (\textit{upper panel}) and change in $M_0$ (\textit{lower panel}) for the trained SALT3.LOWMASS and SALT3.HIGHMASS surfaces. $\Delta M_0 = M_0$(SALT3.HIGHMASS) - $M_0$(SALT3.LOWMASS). The Ca H\&K and Si II features discussed in \citetalias{jones_22_host} are labelled.}
  \label{M0_p3}
\end{figure*}

\begin{figure*}
  \begin{center}
   \includegraphics[width=0.75\textwidth]{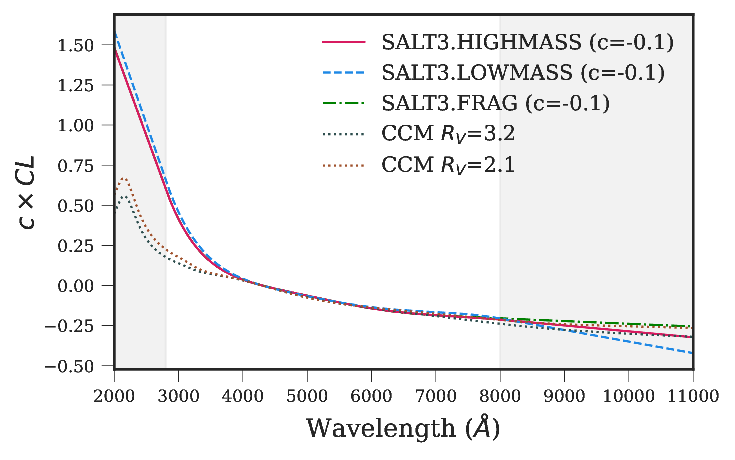}
  \end{center}
  \caption{The colour law component of our trained SALT3.LOWMASS and SALT3.HIGHMASS surfaces, as well as SALT3.FRAG, for a $c=-0.1$ SN. For reference, we also plot the Milky Way laws from \citep{ccmdust} for different values of $R_V$. For much of the wavelength range, the SALT3.HIGHMASS and SALT3.FRAG colour laws directly overlap. The shaded regions indicate the wavelengths where the SALT3 colour law is linearly extrapolated, rather than fit as a polynomial. }
  \label{cl}
\end{figure*}

\subsection{SALT3 Model Components}

The combined $M_0$ and $M_1$ components of SALT3.HIGHMASS and SALT3.LOWMASS at different $x_1$ values are shown in Figure~\ref{sed_ringing}. The difference in the $M_0$ components is shown in Figure~\ref{M0_p3}. These plots show that fiducial SNe~Ia in low-mass galaxies appear to be bluer than those in high-mass galaxies. We observe this trend at all phases. As $M_0$ is by definition the average SED of a $x_1 = 0$, $c=0$ SNe, this result confirms that the definition of $c=0$ changes slightly between the two surfaces.

Apart from the difference in the colour, there are subtle differences in some of the spectral features. The Si~II features in SALT3.HIGHMASS (at 4000, 5770 and 6100\AA) are slightly blueshifted compared to SALT3.LOWMASS. The blueshift reflects the velocity of ejecta coming towards us along the line of sight; higher ejecta velocities are associated with brighter SNe~Ia \citep{benetti05}. The Si~II feature at ~5770\AA is noticeably deeper in SALT3.HIGHMASS. The depth of this line relative to the stronger Si~II feature at 6100\AA correlates with SN~Ia luminosity. Lower ratios correspond to fainter SNe~Ia \citep{nugent_95}. This is inconsistent with the blueshifts seen in the Si~II lines but consistent with the redder colours of SALT3.HIGHMASS.

The $c \times CL$ components (i.e. the exponential term from Equation~\ref{eq:salt_flux_p3}) for a $c=-0.1$ SN are shown in Figure~\ref{cl}. %, highlighting that the a $c=-0.1$ SN will be bluer on average for SALT3.LOWMASS than SALT3.HIGHMASS due to the redefinition of the $c=0$ point. Additionally, 
We find that SNe occurring in low-mass galaxies have a steeper colour law in the UV. % --- so are brighter in the UV than SNe of the same colour from high-mass galaxies. 
\citetalias{jones_22_host} do not train separate phase-independent colour laws for low-host-mass or high-host-mass SNe; but they do find that the phase-\textit{dependent} colours exhibit a change with host-mass. Their results disagree with our findings (though we reiterate that we use different training samples); the \citetalias{jones_22_host} model of high-host-mass SNe is slightly bluer than the low-host-mass model, both at peak phase and in the Lira law tail \citep{lira_98, phillips_99}. They find that these phase-dependent colour differences were previously captured by the $M_1$ component and encompassed by the $x_1$ parameter.

Our results may indicate that SNe in low-mass galaxies are intrinsically more luminous than those in high-mass galaxies. Alternatively, the change in colours between SNe from high- and low-mass hosts may be caused by changes in the extrinsic dust component. %If that is the case, our results suggest that high-mass galaxies have lower total-to-selective dust extinction $R_V$ than their low-mass counterparts, in line with the predictions from \citet{bs20}. 
However, at the current level of modelling it is not possible to disentangle the intrinsic SN colour from the effects of host galaxy dust.

\begin{figure*}
  \begin{center}
    \includegraphics[width=0.9\columnwidth]{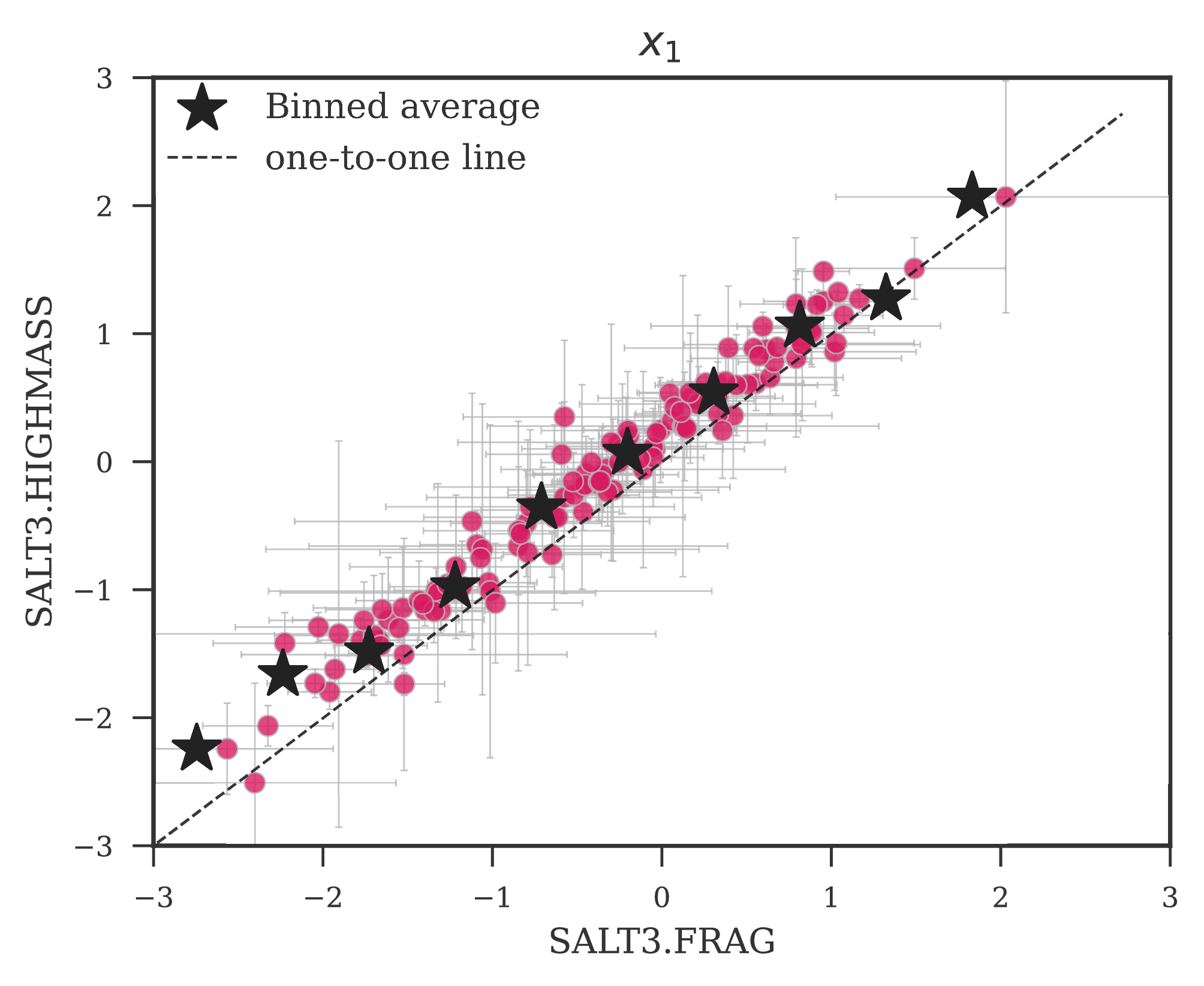}
    \includegraphics[width=0.9\columnwidth]{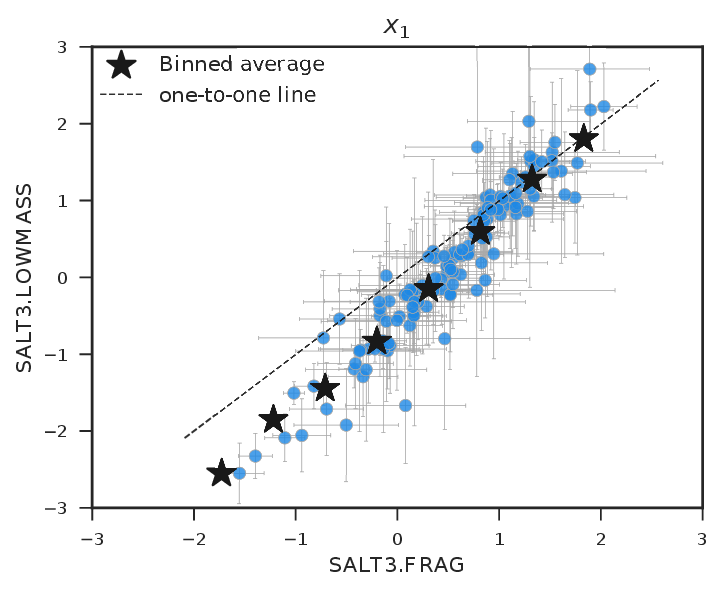}
    \caption{Comparison of the $x_1$ parameter for the DES-SN3YR sample when fit with SALT3.FRAG versus SALT3.HIGHMASS (\textit{left panel, pink points}) and SALT3.LOWMASS (\textit{right panel, blue points}).}
    \label{lcfit_x1}
  \end{center}
\end{figure*}

\begin{figure*}
\centering
   \includegraphics[width=0.9\columnwidth]{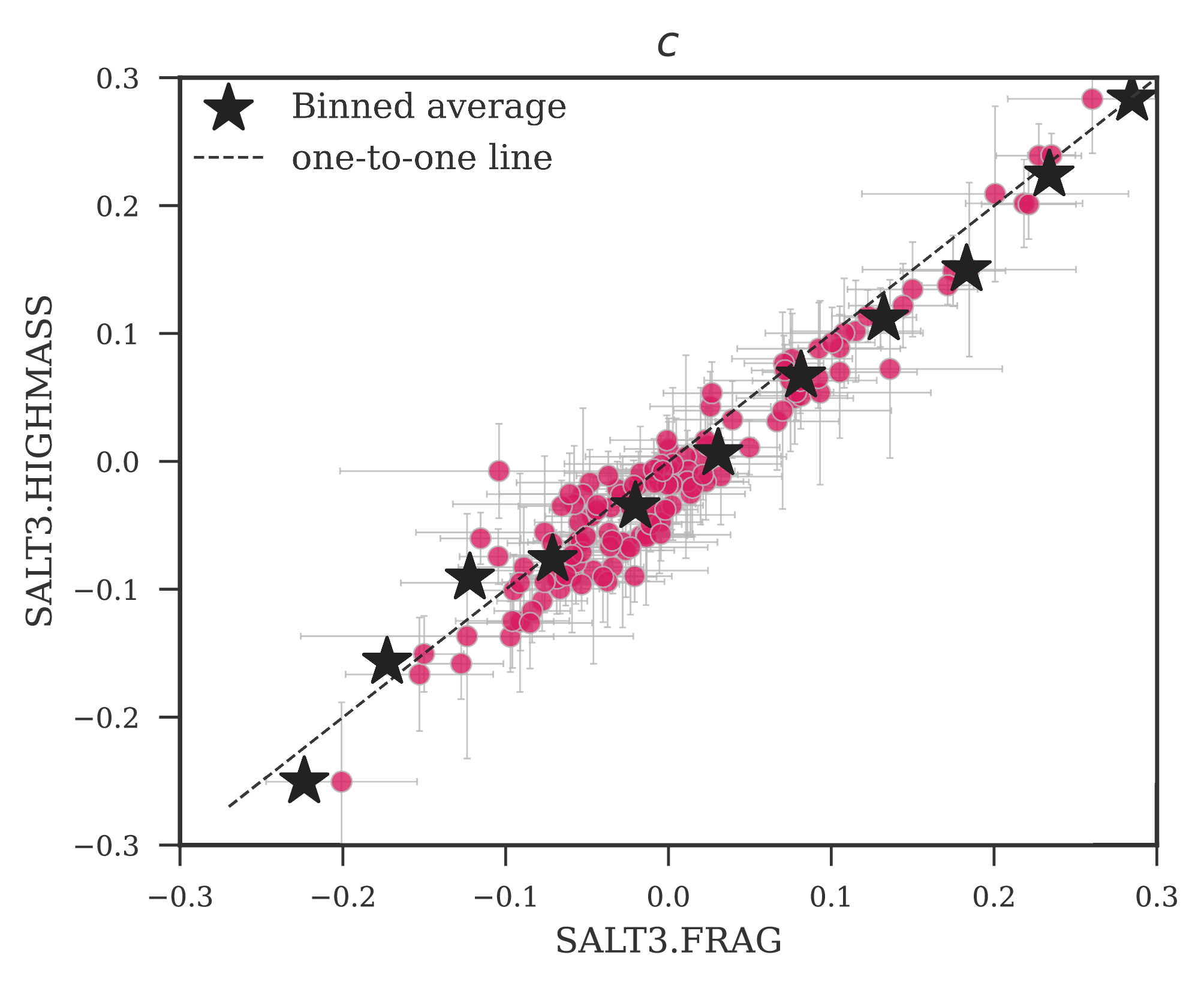}
   \includegraphics[width=0.9\columnwidth]{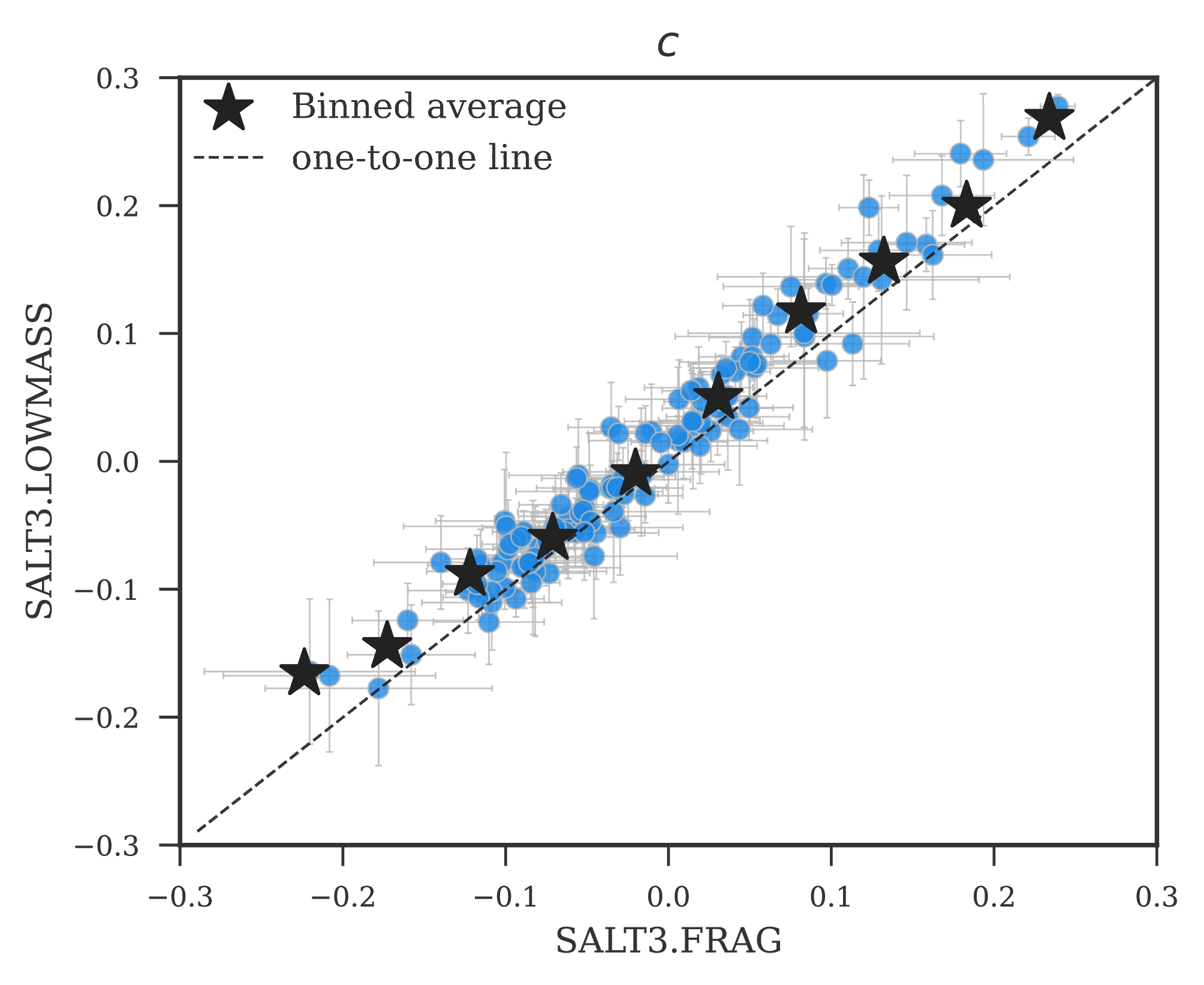}
    \caption{As in Figure~\ref{lcfit_x1}, for $c$.}
    \label{lcfit_c}
\end{figure*}

\begin{figure*}
  \begin{center}
   \includegraphics[width=0.9\columnwidth]{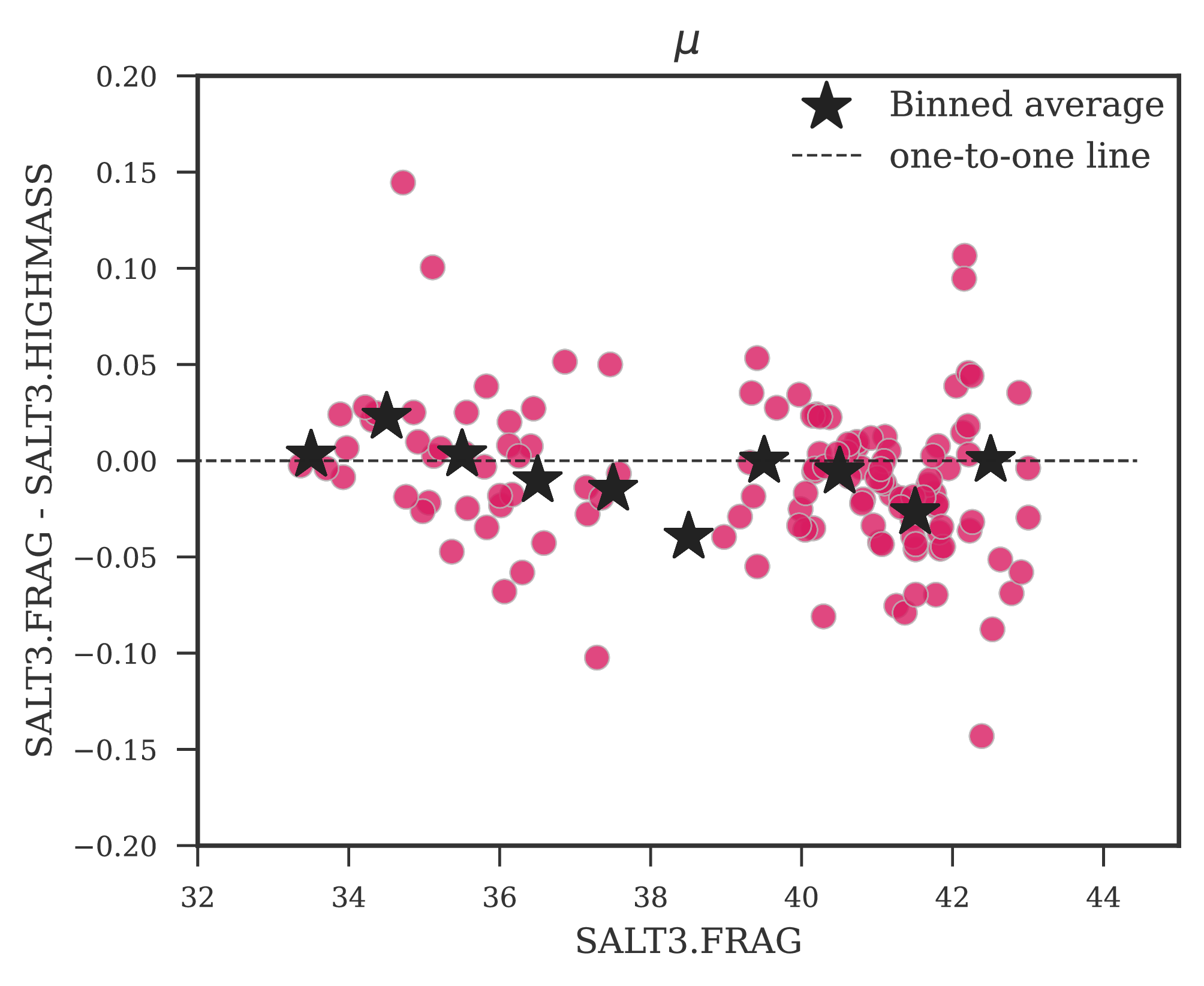}
   \includegraphics[width=0.9\columnwidth]{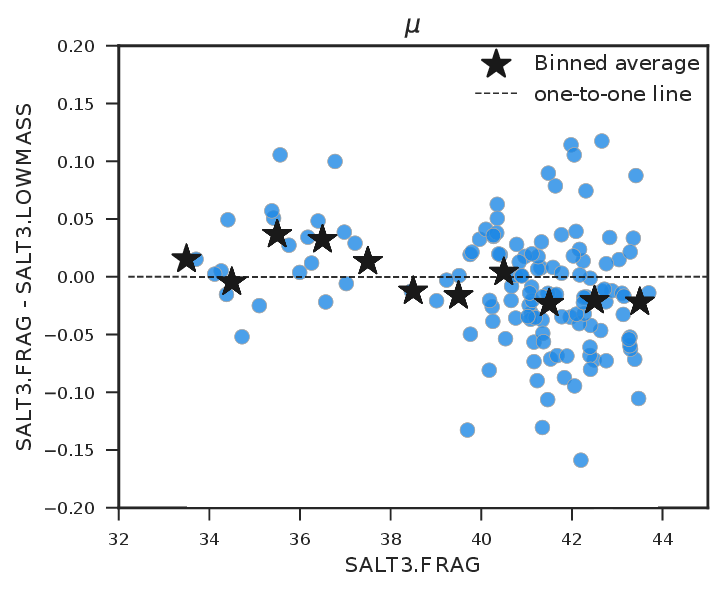}
  \caption{As in Figure~\ref{lcfit_x1}, for distance moduli. We plot the difference in distance moduli on the y-axis.}
  \label{lcfit_mu}
    \end{center}
\end{figure*}

\begin{table*}
\begin{minipage}{\textwidth}
  \renewcommand{\arraystretch}{1.5}
  \centering
  \caption{The best fit nuisance parameters (from Equation~\ref{eq:modified_tripp_p3}) and cosmology parameters for the DES-SN3YR sample when fit with different SALT3 surfaces. All results are obtained using the fast cosmology fitting program, \texttt{wfit}, and include a \citet{planck2015} cosmic microwave background (CMB) prior. As we have used a different set of surfaces to the original DES-SN3YR analysis (which used SALT2.JLA from \citealt{jla}), and apply 1D rather than 5D bias corrections, we do not expect our results to exactly match those presented in \citet{des3yr_systematics, dessnconstraints19}. The SALT3.FRAG (HIGH/LOW) results that are split by host galaxy mass have this selection applied prior to the \texttt{SALT2mu} distance fitting stage. The cosmology fit $\chi^2$ values have been calculated without including contributions from the SALT3 model uncertainties, which change between surfaces.}
    \label{mass_sample_cosmo}
\begin{tabular}{lllllllllll}
\hline
Sample & \# SNe fit & $\alpha$ & $\beta$ & $\sigma_{\rm int}$ & $M$ & $\Omega_M$ & $\sigma_{\Omega_M}^{\rm stat}$ & $w$ & $\sigma_w^{\rm stat}$ & $\chi^2_{\rm cosmo}$ \\ \hline
SALT3.FRAG & 322 & 0.14 $\pm$ 0.01 & 2.78 $\pm$ 0.11 & 0.11 & -29.99 & 0.34 & 0.02 & -0.91 & 0.05 & 11.1 \\
SALT3.FRAG (HIGH) & 187 & 0.15 $\pm$ 0.01 & 2.63 $\pm$ 0.13 & 0.11 & -30.01 & 0.36 & 0.02 & -0.85 & 0.06 & 7.3 \\
SALT3.FRAG (LOW) & 135 & 0.12 $\pm$ 0.02 & 2.99 $\pm$ 0.18 & 0.11 & -29.97 & 0.35 & 0.03 & -0.89 & 0.08 & 12.7 \\
SALT3.FRAG + Mass Step & 322 & 0.15 $\pm$ 0.01 & 2.79 $\pm$ 0.10 & 0.11 & -29.99 & 0.36 & 0.02 & -0.85 & 0.05 & 9.3 \\
SALT3.HIGHMASS & 129 & 0.11 $\pm$ 0.01 & 2.77 $\pm$ 0.16 & 0.09 & -29.94 & 0.33 & 0.02 & -0.96 & 0.07 & 7.9 \\
SALT3.LOWMASS & 134 & 0.06 $\pm$ 0.02 & 2.89 $\pm$ 0.17 & 0.11 & -30.08 & 0.33 & 0.02 & -0.97 & 0.08 & 6.3 \\
\hline
\end{tabular}
\end{minipage}
\end{table*}

\subsection{Light Curve Fits}
As the SN parameters and nuisance parameters are defined on independent scales for each SALT3 surface, we cannot directly compare these results. For example, Figures~\ref{lcfit_x1}-\ref{lcfit_c} show that the definitions of $c$ and $x_1$ have systematically changed between the three surfaces used in our analysis. We find that $x_1({\rm SALT3.LOWMASS})$ is systematically lower than $x_1({\rm SALT3.FRAG})$, and $x_1({\rm SALT3.HIGHMASS})$ is systematically higher. This is consistent with the findings of \citetalias{jones_22_host}, as shown in their Figure~8 (where their $x_1$ parameter is redefined to only include host-mass independent variability). We also find a (weaker) redefinition of $c$ such that for the same SN, $c({\rm SALT3.LOWMASS})$ appears to be systematically higher than $c({\rm SALT3.FRAG})$, and $c({\rm SALT3.HIGHMASS})$ appears to be systematically lower. This follows directly from the $c$ rescaling expected based on Figure~\ref{mass-dist}, in which the mean (i.e. $c_{\rm model}=0$ point) is lower for SALT.LOWMASS and higher for SALT3.HIGHMASS. In contrast, the results from \citetalias{jones_22_host} (where $c$ is defined on the same scale for all models) indicate that in reality, $c({\rm SALT3.LOWMASS})$ is systematically \textit{lower} than $c({\rm SALT3.K21})$, and $c({\rm SALT3.HIGHMASS})$ is slightly higher. This highlights the complications introduced by the floating definitions of $x_1$ and $c$.

Similarly, the nuisance parameters $\alpha$ and $\beta$ (which quantify the global stretch-luminosity and colour-luminosity relationships of SNe~Ia) are defined on different scales for each surface --- though this effect is very slight for $\beta$. The nuisance parameters are presented in Table~\ref{mass_sample_cosmo}. 

The intricacies of comparing fitted light curve parameter values that we have demonstrated here will also apply when comparing any fitted light curve parameters from models trained on different samples (e.g. the Dark Energy Survey's 3-year versus 5-year SN analyses, which use light curve models from \citealt{jla} and \citealt{t23} respectively).

During light curve fitting, we lose four SNe to light curve fitting cuts related to the $x_1$ and $c$ bounds across the LOWMASS and HIGHMASS samples. A further five fail the requirement of having at least one observation before peak $B$-band maximum (as their estimated peak MJD have shifted when fit with our new surfaces). We lose 56 SNe due to the fit probability > 0.01 criterion, the majority of which are from the HIGHMASS low-$z$ sample and have high signal-to-noise ratios.\footnote{The fit probability is the probability of finding an equal or larger light curve data-model $\chi^2$, assuming Gaussian-distributed flux uncertainties.} Our SALT3.HIGHMASS model has lower model uncertainties than SALT3.FRAG or SALT3.LOWMASS, which leads to an increased reduced $\chi^2$ for this  data, thus causing it to fail fit probability cuts. This may be remedied in future work by increasing the number of light curve fit iterations when using models with low levels of uncertainty.

\subsection{Distances and Cosmology}
\label{sec:distance_results}

We compute the distance moduli for each set of light curve and nuisance parameter fits. In Figure~\ref{lcfit_mu}, the distance modulus values do not display any systematic offset between the SALT3.FRAG and SALT3.HIGHMASS surfaces (\textit{left panel}). However, the \textit{right panel} shows a slight slope, such that more distant SNe are measured to be (on average) further away with SALT3.LOWMASS than SALT3.FRAG. This trend is not observed when repeating our analysis over a reduced rest-frame wavelength range of $4000 < \lambda_{\rm eff} < 8000$\AA\ (Figure~\ref{mu_noUV}). This abridged range restricts the use of the less reliable UV region of SALT3.LOWMASS, which is more impactful for distant SNe.

The spread of the Hubble residuals (HR) represents the level of remaining scatter in the standardised SNe~Ia luminosities and measurement errors. We calculate Hubble residuals as $\textrm{HR} _{i} = \mu_{\textrm{fit,}i} - \mu_{\textrm{cosmology,}i}$ for each set of results $i$ obtained from our different SALT3 surfaces. In Table~\ref{cosmo_sigma}, we report the root-mean-squared (RMS, i.e., spread of the Hubble residuals) for each set of results. We also report the weighted RMS. %As the uncertainties are used in the cosmology fitting, it is not straightforward to then include the model uncertainties in our weightings. As an approximation, 
We define the weight of a particular HR to be $w = 1/\sigma_{\mu_{\rm fit}}^2$. We note that the mean measured uncertainty in the distance modulus is lower for the SALT3.HIGHMASS results ($\bar{\sigma}_{\mu_{\rm fit}} = 0.146$ mag) than those from SALT3.FRAG ($\bar{\sigma}_{\mu_{\rm fit}} = 0.156$ mag) or SALT3.FRAG + Mass Step ($\bar{\sigma}_{\mu_{\rm fit}} = 0.155$ mag). The mean uncertainty for the SALT3.LOWMASS results is slightly higher ($\bar{\sigma}_{\mu_{\rm fit}} = 0.157$ mag). For each set of results, we estimate the uncertainty in RMS$_{\rm HR}$ using bootstrapping, resampling the HR values 1000 times and reporting the standard deviation in the resulting distribution.

We split the SALT3.FRAG results into low- and high-host-mass subsamples and calculate the HR results for each, using the best-fit cosmology of the full sample. We also report the HR results for the low- and high-host-mass subsamples fit with SALT3.FRAG (a.k.a., SALT3.FRAG (HIGH/LOW)), where the best-fit cosmology parameters are calculated for each subsample independently.

We find measuring the HR of a subsample of SNe from high-mass hosts lowers the RMS$_{\rm HR}$ by 0.5-1.2$\sigma$ (depending on weighting, application of mass step, and subsample used to calculate best-fit cosmology) compared to using the full DES-SN3YR sample fit with SALT3.FRAG. When fitting the high-host-mass sample with SALT3.HIGHMASS, we find a reduction of 0.1-0.9$\sigma$ in the RMS$_{\rm HR}$ relative to the same sample fit with SALT3.FRAG.

On the other hand, using a subsample of SNe from low-mass hosts \textit{increases} the RMS$_{\rm HR}$ by 0.5-1.2$\sigma$, compared to using the full DES-SN3YR sample fit with SALT3.FRAG, when both samples use the SALT3.FRAG best fit cosmology. When fitting the low-host-mass sample with SALT3.LOWMASS, we intead find a reduction of 0.5-0.7$\sigma$ in the RMS$_{\rm HR}$ relative to the same sample with SALT3.FRAG.

\textbf{These results indicate that the DES-SN3YR SNe in high-mass hosts are better standardisable candles than those in low-mass hosts, and that the standardisability in the DES-SN3YR sample is slightly improved by using custom SALT3.HIGHMASS and SALT3.LOWMASS surfaces (albeit at low statistical significance).} This is consistent with the results of \citetalias{jones_22_host}, who also find negligible changes in the HR scatter.

The SALT3.HIGHMASS and SALT3.LOWMASS surfaces yield similar cosmology results, with the estimated value of $w$ shifting by an insignificant 0.6-1.3$\sigma_w^{\rm stat}$ compared to the SALT3.FRAG results (Table~\ref{mass_sample_cosmo}). Combining the binned Hubble diagrams of the mass-split samples in order to fit a single cosmological model from all the data is not possible with the current fitting infrastructure, but would be necessary if our methods were to be adopted by future cosmology analyses.

\begin{table*}
\begin{minipage}{2\columnwidth}
  \renewcommand{\arraystretch}{1.5}
  \centering
  \caption{Scatter in the Hubble residual (from best-fit cosmology) for each SALT3 surface. All values are in units of magnitude. The best-fit cosmology is unique for each entry in the ``Result" column. The SALT3.FRAG (HIGH) and (LOW) results are split by host galaxy mass prior to the \texttt{SALT2mu} fitting stage, to isolate the effects of fitting cosmology with a subsample of DES-SN3YR SNe versus the effects of changing surfaces.}
    \label{cosmo_sigma}
\begin{tabular}{@{}llll@{}}
\toprule
\rowcolor[HTML]{FFFFFF} 
% \hline
\multicolumn{4}{c}{\textit{Hubble residuals from best-fit cosmology}} \\ \hline
Result & Subsample & RMS$_{\rm HR}$ & Weighted RMS$_{\rm HR}$ \\ \hline
 & Full & 0.155$\pm$0.008 & 0.147$\pm$0.008 \\
 & \cellcolor[HTML]{EFA1BE}$M_{\log} > 10$ & \cellcolor[HTML]{EFA1BE}0.141$\pm$0.009 & \cellcolor[HTML]{EFA1BE}0.137$\pm$0.009 \\
\multirow{-3}{*}{SALT3.FRAG} & \cellcolor[HTML]{90C7F7}$M_{\log} < 10$ & \cellcolor[HTML]{90C7F7}0.173$\pm$0.013 & \cellcolor[HTML]{90C7F7}0.159$\pm$0.014 \\ \hline
 SALT3.FRAG (HIGH) & \cellcolor[HTML]{EFA1BE}$M_{\log} > 10$ & \cellcolor[HTML]{EFA1BE}0.143$\pm$0.009 & \cellcolor[HTML]{EFA1BE}0.139$\pm$0.010 \\
 SALT3.FRAG (LOW) & \cellcolor[HTML]{90C7F7}$M_{\log} < 10$ & \cellcolor[HTML]{90C7F7}0.172$\pm$0.013 & \cellcolor[HTML]{90C7F7}0.156$\pm$0.013 \\ \hline
 & Full & 0.158$\pm$0.008 & 0.149$\pm$0.008 \\
 & \cellcolor[HTML]{EFA1BE}$M_{\log} > 10$ & \cellcolor[HTML]{EFA1BE}0.147$\pm$0.009 & \cellcolor[HTML]{EFA1BE}0.142$\pm$0.010 \\
\multirow{-3}{*}{\begin{tabular}[c]{@{}l@{}}SALT3.FRAG\\  + Mass Step\end{tabular}} & \cellcolor[HTML]{90C7F7}$M_{\log} < 10$ & \cellcolor[HTML]{90C7F7}0.173$\pm$0.013 & \cellcolor[HTML]{90C7F7}0.157$\pm$0.014 \\ \hline
SALT3.HIGHMASS & \cellcolor[HTML]{EFA1BE}$M_{\log} > 10$ & \cellcolor[HTML]{EFA1BE}0.139$\pm$0.011 & \cellcolor[HTML]{EFA1BE}0.128$\pm$0.011 \\ \hline
SALT3.LOWMASS & \cellcolor[HTML]{90C7F7}$M_{\log} < 10$ & \cellcolor[HTML]{90C7F7}0.162$\pm$0.013 & \cellcolor[HTML]{90C7F7}0.146$\pm$0.014 \\ \hline
\end{tabular}
\end{minipage}
\end{table*}

\begin{figure*}
  \begin{center}
   \includegraphics[width=1\textwidth]{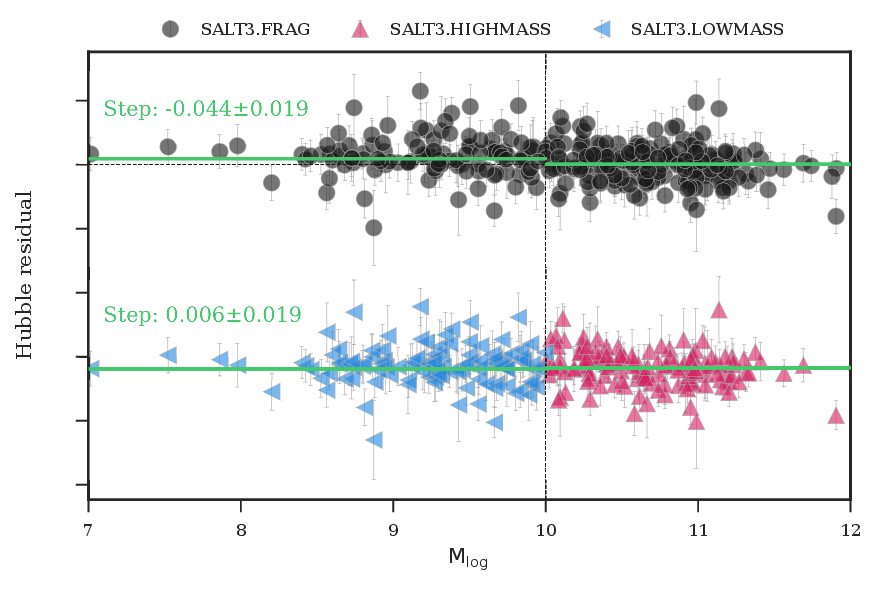}
  \end{center}
  \caption{Correlations between host galaxy masses and Hubble residuals for the DES-SN3YR sample when fit with the SALT3.FRAG (\textit{upper plot}), SALT3.HIGHMASS, and SALT3.LOWMASS surfaces. Here, the Hubble residuals are taken from the best-fit cosmology for each model; a similar result is obtained when calculating Hubble residuals from a constant reference cosmology (i.e., $\Lambda$CDM from \citealt{planck_2018}). The mass step for each set of results is approximated as the difference in weighted mean Hubble residual between the high- and low-host-mass samples, with uncertainties estimated using bootstrapping. We plot a truncated mass range for visual clarity, but use the full range for the mass step calculations. The Hubble residuals are offset on the y-axis for each set of results. In the lower plot, the mass-step is $\sim 0$ by design; the change in luminosity is absorbed by the SALT3 model and the $M$ term of Equation~\ref{eq:modified_tripp_p3} (see Table~\ref{mass_sample_cosmo}).}
  \label{HR_trends_host}
\end{figure*}

\subsection{The Mass Step}
In Figure~\ref{HR_trends_host}, we plot the Hubble residuals as a function of $M_{\log}$ for each surface. We also show the binned weighted average the Hubble residuals in the two mass step bins, and approximate the mass step as the difference in weighted averages between the high- and low-host-mass sets of SNe.\footnote{As we perform independent \texttt{SALT2mu} processes for the low- and high-host-mass samples, we cannot simply fit $\gamma$ to obtain the mass step value.} We bootstrap our samples 1000 times to obtain the uncertainty on the mass step. 

We find that our method of treating SNe from low- and high-mass host galaxies as unique populations in the SALT3 framework removes the observed mass step as efficiently as our explicit fitting for $\gamma$, from step = -0.044 $\pm$ 0.019 mag (SALT3.FRAG) to step = 0.006 $\pm$ 0.019 mag (SALT3.LOWMASS and SALT3.HIGHMASS). However, the fitted fiducial SN ($x_0 = 1, x_1=0, c=0$) absolute magnitude $M$ from Equation~\ref{eq:modified_tripp_p3} changes by a +0.14 magnitudes between SALT3.LOWMASS and SALT3.HIGHMASS, such that SNe in high mass galaxies are fainter. This $\Delta M$ arises from both the redefinition of a fiducial SN between different surfaces \textit{and} the absorption of the host-mass luminosity step. For our SALT3.FRAG (HIGH) and SALT3.FRAG (LOW) results --- which are based on the same surface and so have a consistently defined fiducial SN --- we find $\Delta M = -0.04$. In this case, the observed mass step is also ``removed" (step = -0.0005). This result also holds when calculating HR with respect to a reference $\Lambda$CDM cosmology from \citet{planck_2018}, as was done in \citetalias{jones_22_host}.

Without enforcing a consistent definition of the parameters across different light curve models, we cannot claim that the treatment of the mass-step benefits from adopting bespoke SALT3.HIGHMASS and SALT3.LOWMASS light curve models for sub-populations of SNe in our analysis. The analysis performed by \citetalias{jones_22_host} (in which the parameters are defined on a consistent scale, and the host-mass contribution to the SN flux is modelled and then excluded from the calculation of distances) suggest that improved SALT3 modelling should at least partially reduce the mass-step (by 0.021$\pm$0.002 mag).

\subsection{Future Work}

Here, we have used host-masses from a variety of sources, relying on the broad, binary nature of the observed mass step to mitigate any biases from inconsistent mass estimates. While this has allowed us to use the full constraining power of the \citetalias{salt3} training sample without having to derive our own masses for $\sim 1000$ host galaxies, this analysis should be repeated with more consistently derived masses for both the training and cosmology samples, in order to develop this method into a robust tool for current SN cosmology analyses. For example, we use the DES-SN3YR SVA1-Gold catalogue host-masses described in \citet{des3yr_systematics} for the DES-SN training and cosmology samples, but more recent work by \citet{wiseman_2020, smith2020cosmology} updates these masses using the full 5-yr DES deep-stack photometry. 19 DES SNe change ``mass bins" with the rederived masses. Similarly, 107 of our \citetalias{salt3} training sample are assigned to different mass bins in the Pantheon+ analysis \citep{panplus_datarelease}. Future work should also consider the distributions of host galaxy mass in the training and cosmology samples; many of the SNe used in this work were discovered by targeted surveys which biases our sample. \textbf{Adopting larger training and cosmology samples would reduce the statistical uncertainties on SN distance moduli, providing more clarity on the statistical significance of the potential reduction in HR scatter hinted at in this work.}

An interesting extension to this work would be to apply our methods to other host galaxy parameters, such as the star formation rate. \citet{childress_14} predicts that there are a higher number of young SNe in low-mass galaxies --- and that this relationship appears to evolve with redshift. \citet{childress_14} conclude that SNe~Ia in low-mass, active star-forming galaxies represent a more standardisable population for cosmology analyses. 
We have already seen promising results for SN standardisability when splitting the populations according to host galaxy mass, despite the limitations of our SALT3.LOWMASS model (e.g., inconsistent sources of mass, poor coverage of the training sample, rest-frame UV ringing artefacts). Modelling separate populations of SNe according to parameters such as specific star formation rate (sSFR, i.e. the star formation rate normalised by the stellar mass) may not only help to identify any unique physical processes in the SN~Ia explosions, but reduce the intrinsic scatter remaining in SN cosmology analyses \citep{393, rigault_20}. 

Our straightforward method could be extended to other models that include additional SN~Ia variability (e.g., SNEMO, \citealt{snemo}, SUGAR, \citealt{sugar}, BayeSN \citep{BayeSN}). Using models that include near infra-red data (e.g., BayeSN or SALT3.NIR \citealt{salt3nir}) may be a promising avenue to break the degeneracy between host galaxy mass (or other properties), intrinsic colour, and extrinsic dust effects. However, further use of this method must develop a solution for the pervasive light curve and nuisance parameter scaling problem. \textbf{Without a standardised, surface-independent definition of $x_0, x_1$ and $c$, \textit{any} results derived using different SALT surfaces will be difficult to interpret.} Extensions to this work would also benefit from a program to combine the Hubble diagrams from multiple, independently fit samples into one input for cosmological model fitting (for example, an extension of the current \verb|Pippin| pipeline for SN cosmology, \citealt{Hinton2020}). Without this infrastructure, the method of \citetalias{jones_22_host} is currently the clearest way to incorporate spectro-photometric modelling of SNe~Ia sub-populations into cosmology analyses.

\section{Summary and Conclusions}
\label{sec:summary}

For decades, SN studies have observed relationships between the luminosity of a SN~Ia and its host galaxy properties. These relationships \textit{may} indicate the existence of intrinsically unique sub-populations within the class of SNe~Ia. If that is case, the overall distance measurements and standardisation (i.e. the distance moduli dispersion/Hubble residual scatter) of SNe~Ia could be improved by modelling each intrinsic population separately.

To test this hypothesis, we have produced and tested separate SALT3 light curve models for SNe~Ia that explode in high- or low-mass host galaxies. This straightforward method utilises publicly available training programs and data samples, as an alternative and complement to the method of \citetalias{jones_22_host} (which alters the SALT3 model to include a host galaxy mass component). 

In either method, the resulting high-resolution model spectrum may provide astrophysical clues as to the origins of the observed mass step that would not be obvious when looking at the population photometry or single-object spectra. We find evidence of minor differences in Si~II spectral features between low host-mass and high host-mass samples, in agreement with \citet{foley_11, pan_2015, siebert_2020, pan_2020, dettman_2021}; \citetalias{jones_22_host}. 

While we find that fiducial SNe in low-mass hosts are intrinsically bluer than those in high-mass hosts, \citetalias{jones_22_host} find the opposite. The inconsistency in the fiducial SN SED (which is defined relative to the mean $x_0, x_1, c$ values of a SALT3 training sample) highlights the sensitivity of the SALT3 model to the training sample demographics. This simple finding reinforces the importance of a quality, unbiased light curve training sample (which must be comprehensive and representative of the natural SN populations) for SN Ia analyses \citep{dai_22}.

Here, the fitted SN parameters ${x_0, x_1, c}$ are defined independently for each surface, leading to unique definitions of the fitted nuisance parameters $m_B$, $\alpha$, $\beta$ and $M$ for each surface. We assume that the standardised distance moduli are defined consistently across all surfaces, and so can be fairly compared. We recover consistent cosmological parameters for a flat $w$CDM model (with CMB priors from \citealt{planck2015}) from our low host-mass and high host-mass analyses. We find that fitting the high host-mass subsample of SNe~Ia from DES-SN3YR is intrinsically more standardisable than the low host-mass subsample when fit with the same SALT3.FRAG surface, with a $\sim 2\sigma$ difference between the Hubble residual RMS of these subsamples. Using our custom SALT3.LOWMASS and SALT3.HIGHMASS surfaces to fit these subsamples reduces the Hubble residual RMS of both by $<1\sigma$. Though \citetalias{jones_22_host} suggest that the observed mass-step in the Hubble residuals should be reduced when including host mass considerations in the SALT3 modelling, the varying definitions of the nuisance parameter $M$ mean we cannot conclusively claim this is the case in our analysis. Standardising the definitions of $x_0, x_1$ and $c$ (and therefore, the nuisance parameters of Equation~\ref{eq:modified_tripp_p3}) across different light curve models is a critical step in realising the full potential of our method. This is also important for comparing \textit{any} results generated using different light curve models; the nuance of the redefined parameters may not be obvious, but it can be impactful.

We have demonstrated a novel application of the SALT3 light curve model training procedure, that has the potential to improve SN~Ia standardisability. Our straightforward method may be easily used to incorporate new, high-quality training data or explore other host-galaxy relationships. Moreover, this is proof of how far the light curve model training infrastructure has developed in recent years: the \verb|SALTShaker| training code we use is publicly available, well documented (and tested), and actively maintained, making it a useful tool for the entire SN community.

\section*{Data Availability}
All data and training software underlying this work is publicly available.

\section*{Acknowledgements}
This research was completed in part on Ngunnawal, Ngambri, and Whadjuk Nyoongar country. \\
We thank the SALT3 development team, particularly David Jones and D'Arcy Kenworthy, for their helpful discussions. We also thank Mark Sullivan for providing us with the SNLS host mass data. This research was supported by an Australian Government Research Training Program (RTP) Scholarship. This project used public archival data from the Dark Energy Survey (DES). Funding for the DES Projects has been provided by the U.S. Department of Energy, the U.S. National Science Foundation, the Ministry of Science and Education of Spain, the Science and Technology Facilities Council of the United Kingdom, the Higher Education Funding Council for England, the National Center for Supercomputing Applications at the University of Illinois at Urbana-Champaign, the Kavli Institute of Cosmological Physics at the University of Chicago, the Center for Cosmology and Astro-Particle Physics at the Ohio State University, the Mitchell Institute for Fundamental Physics and Astronomy at Texas A\&M University, Financiadora de Estudos e Projetos, Funda{\c c}{\~a}o Carlos Chagas Filho de Amparo {\`a} Pesquisa do Estado do Rio de Janeiro, Conselho Nacional de Desenvolvimento Cient{\'i}fico e Tecnol{\'o}gico and the Minist{\'e}rio da Ci{\^e}ncia, Tecnologia e Inova{\c c}{\~a}o, the Deutsche Forschungsgemeinschaft, and the Collaborating Institutions in the Dark Energy Survey. The Collaborating Institutions are Argonne National Laboratory, the University of California at Santa Cruz, the University of Cambridge, Centro de Investigaciones Energ{\'e}ticas, Medioambientales y Tecnol{\'o}gicas-Madrid, the University of Chicago, University College London, the DES-Brazil Consortium, the University of Edinburgh, the Eidgen{\"o}ssische Technische Hochschule (ETH) Z{\"u}rich,  Fermi National Accelerator Laboratory, the University of Illinois at Urbana-Champaign, the Institut de Ci{\`e}ncies de l'Espai (IEEC/CSIC), the Institut de F{\'i}sica d'Altes Energies, Lawrence Berkeley National Laboratory, the Ludwig-Maximilians Universit{\"a}t M{\"u}nchen and the associated Excellence Cluster Universe, the University of Michigan, the National Optical Astronomy Observatory, the University of Nottingham, The Ohio State University, the OzDES Membership Consortium, the University of Pennsylvania, the University of Portsmouth, SLAC National Accelerator Laboratory, Stanford University, the University of Sussex, and Texas A\&M University.
Based in part on observations at Cerro Tololo Inter-American Observatory, National Optical Astronomy Observatory, which is operated by the Association of Universities for Research in Astronomy (AURA) under a cooperative agreement with the National Science Foundation.

%%%%%%%%%%%%%%%%%%%%%%%%%%%%%%%%%%%%%%%%%%%%%%%%%%



\bibliographystyle{mnras}
\bibliography{refs} % if your bibtex file is called example.bib

\begin{thebibliography}{}
\makeatletter
\relax
\def\mn@urlcharsother{\let\do\@makeother \do\$\do\&\do\#\do\^\do\_\do\%\do\~}
\def\mn@doi{\begingroup\mn@urlcharsother \@ifnextchar [ {\mn@doi@}
  {\mn@doi@[]}}
\def\mn@doi@[#1]#2{\def\@tempa{#1}\ifx\@tempa\@empty \href
  {http://dx.doi.org/#2} {doi:#2}\else \href {http://dx.doi.org/#2} {#1}\fi
  \endgroup}
\def\mn@eprint#1#2{\mn@eprint@#1:#2::\@nil}
\def\mn@eprint@arXiv#1{\href {http://arxiv.org/abs/#1} {{\tt arXiv:#1}}}
\def\mn@eprint@dblp#1{\href {http://dblp.uni-trier.de/rec/bibtex/#1.xml}
  {dblp:#1}}
\def\mn@eprint@#1:#2:#3:#4\@nil{\def\@tempa {#1}\def\@tempb {#2}\def\@tempc
  {#3}\ifx \@tempc \@empty \let \@tempc \@tempb \let \@tempb \@tempa \fi \ifx
  \@tempb \@empty \def\@tempb {arXiv}\fi \@ifundefined
  {mn@eprint@\@tempb}{\@tempb:\@tempc}{\expandafter \expandafter \csname
  mn@eprint@\@tempb\endcsname \expandafter{\@tempc}}}

\bibitem[\protect\citeauthoryear{{Abbott} et~al.,}{{Abbott}
  et~al.}{2019}]{dessnconstraints19}
{Abbott} T.~M.~C.,  et~al., 2019, \mn@doi [\apj] {10.3847/2041-8213/ab04fa},
  \href {https://ui.adsabs.harvard.edu/abs/2019ApJ...872L..30A} {872, L30}

\bibitem[\protect\citeauthoryear{{Astier} et~al.,}{{Astier}
  et~al.}{2006}]{snls06}
{Astier} P.,  et~al., 2006, \mn@doi [\aap] {10.1051/0004-6361:20054185}, \href
  {http://adsabs.harvard.edu/abs/2006A\%26A...447...31A} {447, 31}

\bibitem[\protect\citeauthoryear{Benetti et~al.,}{Benetti
  et~al.}{2005}]{benetti05}
Benetti S.,  et~al., 2005, \mn@doi [The Astrophysical Journal]
  {10.1086/428608}, 623, 1011

\bibitem[\protect\citeauthoryear{{Betoule} et~al.,}{{Betoule}
  et~al.}{2014}]{jla}
{Betoule} M.,  et~al., 2014, \mn@doi [\aap] {10.1051/0004-6361/201423413},
  \href {http://adsabs.harvard.edu/abs/2014A\%26A...568A..22B} {568, A22}

\bibitem[\protect\citeauthoryear{{Branch}}{{Branch}}{1998}]{branch_98_h0}
{Branch} D.,  1998, \mn@doi [\araa] {10.1146/annurev.astro.36.1.17}, \href
  {https://ui.adsabs.harvard.edu/abs/1998ARA&A..36...17B} {36, 17}

\bibitem[\protect\citeauthoryear{{Branch}, {Romanishin}  \& {Baron}}{{Branch}
  et~al.}{1996}]{branch_96}
{Branch} D.,  {Romanishin} W.,   {Baron} E.,  1996, \mn@doi [\apj]
  {10.1086/177402}, \href
  {https://ui.adsabs.harvard.edu/abs/1996ApJ...465...73B} {465, 73}

\bibitem[\protect\citeauthoryear{{Briday} et~al.,}{{Briday}
  et~al.}{2022}]{briday_22}
{Briday} M.,  et~al., 2022, \mn@doi [\aap] {10.1051/0004-6361/202141160}, \href
  {https://ui.adsabs.harvard.edu/abs/2022A&A...657A..22B} {657, A22}

\bibitem[\protect\citeauthoryear{Brout \& Scolnic}{Brout \&
  Scolnic}{2021}]{bs20}
Brout D.,  Scolnic D.,  2021, \mn@doi [The Astrophysical Journal]
  {10.3847/1538-4357/abd69b}, 909, 26

\bibitem[\protect\citeauthoryear{Brout et~al.,}{Brout et~al.}{2019a}]{deslc}
Brout D.,  et~al., 2019a, \mn@doi [The Astrophysical Journal]
  {10.3847/1538-4357/ab06c1}, 874, 106

\bibitem[\protect\citeauthoryear{{Brout} et~al.,}{{Brout}
  et~al.}{2019b}]{des3yr_systematics}
{Brout} D.,  et~al., 2019b, \mn@doi [\apj] {10.3847/1538-4357/ab08a0}, \href
  {https://ui.adsabs.harvard.edu/abs/2019ApJ...874..150B} {874, 150}

\bibitem[\protect\citeauthoryear{Brout et~al.,}{Brout
  et~al.}{2022a}]{panplus_constraints}
Brout D.,  et~al., 2022a, The Pantheon+ Analysis: Cosmological Constraints,
  \mn@doi{10.48550/ARXIV.2202.04077}, \url {https://arxiv.org/abs/2202.04077}

\bibitem[\protect\citeauthoryear{{Brout} et~al.,}{{Brout}
  et~al.}{2022b}]{superfrag}
{Brout} D.,  et~al., 2022b, \mn@doi [\apj] {10.3847/1538-4357/ac8bcc}, \href
  {https://ui.adsabs.harvard.edu/abs/2022ApJ...938..111B} {938, 111}

\bibitem[\protect\citeauthoryear{{Cardelli}, {Clayton}  \& {Mathis}}{{Cardelli}
  et~al.}{1989}]{ccmdust}
{Cardelli} J.~A.,  {Clayton} G.~C.,   {Mathis} J.~S.,  1989, \mn@doi [\apj]
  {10.1086/167900}, \href {http://adsabs.harvard.edu/abs/1989ApJ...345..245C}
  {345, 245}

\bibitem[\protect\citeauthoryear{{Childress}, {Wolf}  \& {Zahid}}{{Childress}
  et~al.}{2014}]{childress_14}
{Childress} M.~J.,  {Wolf} C.,   {Zahid} H.~J.,  2014, \mn@doi [\mnras]
  {10.1093/mnras/stu1892}, \href
  {https://ui.adsabs.harvard.edu/abs/2014MNRAS.445.1898C} {445, 1898}

\bibitem[\protect\citeauthoryear{{Conley} et~al.,}{{Conley}
  et~al.}{2011}]{conley11}
{Conley} A.,  et~al., 2011, \mn@doi [\apjs] {10.1088/0067-0049/192/1/1}, \href
  {http://adsabs.harvard.edu/abs/2011ApJS..192....1C} {192, 1}

\bibitem[\protect\citeauthoryear{D{\textquotesingle}Andrea
  et~al.,}{D{\textquotesingle}Andrea et~al.}{2011}]{D_Andrea_2011}
D{\textquotesingle}Andrea C.~B.,  et~al., 2011, \mn@doi [The Astrophysical
  Journal] {10.1088/0004-637x/743/2/172}, 743, 172

\bibitem[\protect\citeauthoryear{Dai, Jones, Kenworthy, Kessler, Pierel, Foley,
  Jha  \& Scolnic}{Dai et~al.}{2022}]{dai_22}
Dai M.,  Jones D.~O.,  Kenworthy W.~D.,  Kessler R.,  Pierel J. D.~R.,  Foley
  R.~J.,  Jha S.~W.,   Scolnic D.~M.,  2022, Propagating Uncertainties in the
  SALT3 Model Training Process to Cosmological Constraints,
  \mn@doi{10.48550/ARXIV.2212.06879}, \url {https://arxiv.org/abs/2212.06879}

\bibitem[\protect\citeauthoryear{{Dettman} et~al.,}{{Dettman}
  et~al.}{2021}]{dettman_2021}
{Dettman} K.~G.,  et~al., 2021, \mn@doi [\apj] {10.3847/1538-4357/ac2ee5},
  \href {https://ui.adsabs.harvard.edu/abs/2021ApJ...923..267D} {923, 267}

\bibitem[\protect\citeauthoryear{{Foley} \& {Kasen}}{{Foley} \&
  {Kasen}}{2011}]{foley_11}
{Foley} R.~J.,  {Kasen} D.,  2011, \mn@doi [\apj] {10.1088/0004-637X/729/1/55},
  \href {https://ui.adsabs.harvard.edu/abs/2011ApJ...729...55F} {729, 55}

\bibitem[\protect\citeauthoryear{{Galbany} et~al.,}{{Galbany}
  et~al.}{2022}]{galbany_22}
{Galbany} L.,  et~al., 2022, \mn@doi [arXiv e-prints]
  {10.48550/arXiv.2209.02546}, \href
  {https://ui.adsabs.harvard.edu/abs/2022arXiv220902546G} {p. arXiv:2209.02546}

\bibitem[\protect\citeauthoryear{{Gallagher}, {Garnavich}, {Berlind},
  {Challis}, {Jha}  \& {Kirshner}}{{Gallagher} et~al.}{2005}]{gallagher_2005}
{Gallagher} J.~S.,  {Garnavich} P.~M.,  {Berlind} P.,  {Challis} P.,  {Jha} S.,
    {Kirshner} R.~P.,  2005, \mn@doi [\apj] {10.1086/491664}, \href
  {https://ui.adsabs.harvard.edu/abs/2005ApJ...634..210G} {634, 210}

\bibitem[\protect\citeauthoryear{Graur et~al.,}{Graur
  et~al.}{2014}]{Graur_2014_highzSN}
Graur O.,  et~al., 2014, \mn@doi [The Astrophysical Journal]
  {10.1088/0004-637x/783/1/28}, 783, 28

\bibitem[\protect\citeauthoryear{{Guy} et~al.,}{{Guy} et~al.}{2007}]{salt2}
{Guy} J.,  et~al., 2007, \mn@doi [\aap] {10.1051/0004-6361:20066930}, \href
  {http://adsabs.harvard.edu/abs/2007A\%26A...466...11G} {466, 11}

\bibitem[\protect\citeauthoryear{{Guy} et~al.,}{{Guy}
  et~al.}{2010}]{g10_scatter}
{Guy} J.,  et~al., 2010, \mn@doi [\aap] {10.1051/0004-6361/201014468}, \href
  {http://adsabs.harvard.edu/abs/2010A\%26A...523A...7G} {523, A7}

\bibitem[\protect\citeauthoryear{{Hamuy}, {Phillips}, {Maza}, {Suntzeff},
  {Schommer}  \& {Aviles}}{{Hamuy} et~al.}{1995}]{hamuy_95}
{Hamuy} M.,  {Phillips} M.~M.,  {Maza} J.,  {Suntzeff} N.~B.,  {Schommer}
  R.~A.,   {Aviles} R.,  1995, \mn@doi [\aj] {10.1086/117251}, \href
  {https://ui.adsabs.harvard.edu/abs/1995AJ....109....1H} {109, 1}

\bibitem[\protect\citeauthoryear{{Hamuy}, {Trager}, {Pinto}, {Phillips},
  {Schommer}, {Ivanov}  \& {Suntzeff}}{{Hamuy} et~al.}{2000}]{hamuy00}
{Hamuy} M.,  {Trager} S.~C.,  {Pinto} P.~A.,  {Phillips} M.~M.,  {Schommer}
  R.~A.,  {Ivanov} V.,   {Suntzeff} N.~B.,  2000, \mn@doi [\aj]
  {10.1086/301527}, \href
  {https://ui.adsabs.harvard.edu/abs/2000AJ....120.1479H} {120, 1479}

\bibitem[\protect\citeauthoryear{{Hand}, {Liu}, {Galbany}, {Perrefort},
  {Wood-Vasey}  \& {Burns}}{{Hand} et~al.}{2022}]{hand_22}
{Hand} J.,  {Liu} S.,  {Galbany} L.,  {Perrefort} D.,  {Wood-Vasey} W.~M.,
  {Burns} C.,  2022, \mn@doi [\apj] {10.3847/1538-4357/ac389f}, \href
  {https://ui.adsabs.harvard.edu/abs/2022ApJ...925..115H} {925, 115}

\bibitem[\protect\citeauthoryear{{Hayden}, {Gupta}, {Garnavich}, {Mannucci},
  {Nichol}  \& {Sako}}{{Hayden} et~al.}{2013}]{hayden_13}
{Hayden} B.~T.,  {Gupta} R.~R.,  {Garnavich} P.~M.,  {Mannucci} F.,  {Nichol}
  R.~C.,   {Sako} M.,  2013, \mn@doi [\apj] {10.1088/0004-637X/764/2/191},
  \href {https://ui.adsabs.harvard.edu/abs/2013ApJ...764..191H} {764, 191}

\bibitem[\protect\citeauthoryear{{Hayden} et~al.,}{{Hayden}
  et~al.}{2021}]{hayden_2021}
{Hayden} B.,  et~al., 2021, \mn@doi [\apj] {10.3847/1538-4357/abed4d}, \href
  {https://ui.adsabs.harvard.edu/abs/2021ApJ...912...87H} {912, 87}

\bibitem[\protect\citeauthoryear{Hinton \& Brout}{Hinton \&
  Brout}{2020}]{Hinton2020}
Hinton S.,  Brout D.,  2020, \mn@doi [Journal of Open Source Software]
  {10.21105/joss.02122}, 5, 2122

\bibitem[\protect\citeauthoryear{{Holtzman} et~al.,}{{Holtzman}
  et~al.}{2008}]{holtzman_08}
{Holtzman} J.~A.,  et~al., 2008, \mn@doi [\aj] {10.1088/0004-6256/136/6/2306},
  \href {https://ui.adsabs.harvard.edu/abs/2008AJ....136.2306H} {136, 2306}

\bibitem[\protect\citeauthoryear{{Howell}}{{Howell}}{2001}]{howell_01}
{Howell} D.~A.,  2001, \mn@doi [\apjl] {10.1086/321702}, \href
  {https://ui.adsabs.harvard.edu/abs/2001ApJ...554L.193H} {554, L193}

\bibitem[\protect\citeauthoryear{{Jha} et~al.,}{{Jha} et~al.}{2006}]{jha_06}
{Jha} S.,  et~al., 2006, \mn@doi [\aj] {10.1086/497989}, \href
  {https://ui.adsabs.harvard.edu/abs/2006AJ....131..527J} {131, 527}

\bibitem[\protect\citeauthoryear{{Johansson} et~al.,}{{Johansson}
  et~al.}{2021}]{johansson_21}
{Johansson} J.,  et~al., 2021, \mn@doi [\apj] {10.3847/1538-4357/ac2f9e}, \href
  {https://ui.adsabs.harvard.edu/abs/2021ApJ...923..237J} {923, 237}

\bibitem[\protect\citeauthoryear{Jones \& Kessler}{Jones \&
  Kessler}{2018}]{saltshaker_doc}
Jones D.,  Kessler R.,  2018, Saltshaker

\bibitem[\protect\citeauthoryear{Jones et~al.,}{Jones
  et~al.}{2013}]{Jones_2013_highz}
Jones D.~O.,  et~al., 2013, \mn@doi [The Astrophysical Journal]
  {10.1088/0004-637x/768/2/166}, 768, 166

\bibitem[\protect\citeauthoryear{{Jones} et~al.,}{{Jones}
  et~al.}{2018}]{jones-18_local}
{Jones} D.~O.,  et~al., 2018, \mn@doi [\apj] {10.3847/1538-4357/aae2b9}, \href
  {https://ui.adsabs.harvard.edu/abs/2018ApJ...867..108J} {867, 108}

\bibitem[\protect\citeauthoryear{Jones et~al.,}{Jones
  et~al.}{2019}]{Jones_2019}
Jones D.~O.,  et~al., 2019, \mn@doi [The Astrophysical Journal]
  {10.3847/1538-4357/ab2bec}, 881, 19

\bibitem[\protect\citeauthoryear{{Jones}, {Kenworthy}, {Dai}, {Foley},
  {Kessler}, {Pierel}  \& {Siebert}}{{Jones} et~al.}{2023}]{jones_22_host}
{Jones} D.~O.,  {Kenworthy} W.~D.,  {Dai} M.,  {Foley} R.~J.,  {Kessler} R.,
  {Pierel} J.~D.~R.,   {Siebert} M.~R.,  2023, \mn@doi [\apj]
  {10.3847/1538-4357/acd195}, \href
  {https://ui.adsabs.harvard.edu/abs/2023ApJ...951...22J} {951, 22}

\bibitem[\protect\citeauthoryear{{Kelly}, {Hicken}, {Burke}, {Mand el}  \&
  {Kirshner}}{{Kelly} et~al.}{2010}]{kelly2010}
{Kelly} P.~L.,  {Hicken} M.,  {Burke} D.~L.,  {Mand el} K.~S.,   {Kirshner}
  R.~P.,  2010, \mn@doi [\apj] {10.1088/0004-637X/715/2/743}, \href
  {https://ui.adsabs.harvard.edu/abs/2010ApJ...715..743K} {715, 743}

\bibitem[\protect\citeauthoryear{{Kelsey} et~al.,}{{Kelsey}
  et~al.}{2021}]{kelsey_2021}
{Kelsey} L.,  et~al., 2021, \mn@doi [\mnras] {10.1093/mnras/staa3924}, \href
  {https://ui.adsabs.harvard.edu/abs/2021MNRAS.501.4861K} {501, 4861}

\bibitem[\protect\citeauthoryear{{Kelsey} et~al.,}{{Kelsey}
  et~al.}{2023}]{kelsey_23}
{Kelsey} L.,  et~al., 2023, \mn@doi [\mnras] {10.1093/mnras/stac3711}, \href
  {https://ui.adsabs.harvard.edu/abs/2023MNRAS.519.3046K} {519, 3046}

\bibitem[\protect\citeauthoryear{{Kenworthy}}{{Kenworthy}}{2023}]{kenworthy_23}
{Kenworthy} W.~D. e.~a.,  in prep. 2023

\bibitem[\protect\citeauthoryear{Kenworthy et~al.,}{Kenworthy
  et~al.}{2021}]{salt3}
Kenworthy W.~D.,  et~al., 2021, SALT3: An Improved Type Ia Supernova Model for
  Measuring Cosmic Distances (\mn@eprint {arXiv} {2104.07795})

\bibitem[\protect\citeauthoryear{{Kessler} et~al.,}{{Kessler}
  et~al.}{2009}]{snana09}
{Kessler} R.,  et~al., 2009, \mn@doi [\pasp] {10.1086/605984}, \href
  {http://adsabs.harvard.edu/abs/2009PASP..121.1028K} {121, 1028}

\bibitem[\protect\citeauthoryear{Kessler et~al.,}{Kessler
  et~al.}{2013}]{Kessler_2013}
Kessler R.,  et~al., 2013, \mn@doi [The Astrophysical Journal]
  {10.1088/0004-637x/764/1/48}, 764, 48

\bibitem[\protect\citeauthoryear{Kessler et~al.,}{Kessler
  et~al.}{2019}]{Kessler_2019_sims}
Kessler R.,  et~al., 2019, \mn@doi [Monthly Notices of the Royal Astronomical
  Society] {10.1093/mnras/stz463}, 485, 1171–1187

\bibitem[\protect\citeauthoryear{{Kim}, {Smith}, {Sullivan}  \& {Lee}}{{Kim}
  et~al.}{2018}]{kim_18}
{Kim} Y.-L.,  {Smith} M.,  {Sullivan} M.,   {Lee} Y.-W.,  2018, \mn@doi [\apj]
  {10.3847/1538-4357/aaa127}, \href
  {https://ui.adsabs.harvard.edu/abs/2018ApJ...854...24K} {854, 24}

\bibitem[\protect\citeauthoryear{{Lampeitl} et~al.,}{{Lampeitl}
  et~al.}{2010}]{lampeitl_2010}
{Lampeitl} H.,  et~al., 2010, \mn@doi [\apj] {10.1088/0004-637X/722/1/566},
  \href {https://ui.adsabs.harvard.edu/abs/2010ApJ...722..566L} {722, 566}

\bibitem[\protect\citeauthoryear{L{\'e}get et~al.,}{L{\'e}get
  et~al.}{2019}]{sugar}
L{\'e}get P.~F.,  et~al., 2019, SUGAR: An improved empirical model of Type Ia
  Supernovae based on spectral features (\mn@eprint {arXiv} {1909.11239})

\bibitem[\protect\citeauthoryear{{Lira} et~al.,}{{Lira} et~al.}{1998}]{lira_98}
{Lira} P.,  et~al., 1998, \mn@doi [\aj] {10.1086/300175}, \href
  {https://ui.adsabs.harvard.edu/abs/1998AJ....115..234L} {115, 234}

\bibitem[\protect\citeauthoryear{{Mandel}, {Wood-Vasey}, {Friedman}  \&
  {Kirshner}}{{Mandel} et~al.}{2009}]{mandel_09}
{Mandel} K.~S.,  {Wood-Vasey} W.~M.,  {Friedman} A.~S.,   {Kirshner} R.~P.,
  2009, \mn@doi [\apj] {10.1088/0004-637X/704/1/629}, \href
  {https://ui.adsabs.harvard.edu/abs/2009ApJ...704..629M} {704, 629}

\bibitem[\protect\citeauthoryear{Mandel, Thorp, Narayan, Friedman  \&
  Avelino}{Mandel et~al.}{2020}]{BayeSN}
Mandel K.~S.,  Thorp S.,  Narayan G.,  Friedman A.~S.,   Avelino A.,  2020, A
  Hierarchical Bayesian SED Model for Type Ia Supernovae in the Optical to
  Near-Infrared (\mn@eprint {arXiv} {2008.07538})

\bibitem[\protect\citeauthoryear{{Marriner} et~al.,}{{Marriner}
  et~al.}{2011}]{salt2mu}
{Marriner} J.,  et~al., 2011, \mn@doi [\apj] {10.1088/0004-637X/740/2/72},
  \href {https://ui.adsabs.harvard.edu/abs/2011ApJ...740...72M} {740, 72}

\bibitem[\protect\citeauthoryear{{Mosher} et~al.,}{{Mosher}
  et~al.}{2014}]{mosher14}
{Mosher} J.,  et~al., 2014, \mn@doi [\apj] {10.1088/0004-637X/793/1/16}, \href
  {http://adsabs.harvard.edu/abs/2014ApJ...793...16M} {793, 16}

\bibitem[\protect\citeauthoryear{{Nugent}, {Phillips}, {Baron}, {Branch}  \&
  {Hauschildt}}{{Nugent} et~al.}{1995}]{nugent_95}
{Nugent} P.,  {Phillips} M.,  {Baron} E.,  {Branch} D.,   {Hauschildt} P.,
  1995, \mn@doi [\apjl] {10.1086/309846}, \href
  {https://ui.adsabs.harvard.edu/abs/1995ApJ...455L.147N} {455, L147}

\bibitem[\protect\citeauthoryear{{Pan}}{{Pan}}{2020}]{pan_2020}
{Pan} Y.-C.,  2020, \mn@doi [\apjl] {10.3847/2041-8213/ab8e47}, \href
  {https://ui.adsabs.harvard.edu/abs/2020ApJ...895L...5P} {895, L5}

\bibitem[\protect\citeauthoryear{{Pan}, {Sullivan}, {Maguire}, {Gal-Yam},
  {Hook}, {Howell}, {Nugent}  \& {Mazzali}}{{Pan} et~al.}{2015}]{pan_2015}
{Pan} Y.~C.,  {Sullivan} M.,  {Maguire} K.,  {Gal-Yam} A.,  {Hook} I.~M.,
  {Howell} D.~A.,  {Nugent} P.~E.,   {Mazzali} P.~A.,  2015, \mn@doi [\mnras]
  {10.1093/mnras/stu2121}, \href
  {https://ui.adsabs.harvard.edu/abs/2015MNRAS.446..354P} {446, 354}

\bibitem[\protect\citeauthoryear{{Phillips}, {Lira}, {Suntzeff}, {Schommer},
  {Hamuy}  \& {Maza}}{{Phillips} et~al.}{1999}]{phillips_99}
{Phillips} M.~M.,  {Lira} P.,  {Suntzeff} N.~B.,  {Schommer} R.~A.,  {Hamuy}
  M.,   {Maza} J.,  1999, \mn@doi [\aj] {10.1086/301032}, \href
  {https://ui.adsabs.harvard.edu/abs/1999AJ....118.1766P} {118, 1766}

\bibitem[\protect\citeauthoryear{Pierel et~al.,}{Pierel
  et~al.}{2022}]{salt3nir}
Pierel J. D.~R.,  et~al., 2022, SALT3-NIR: Taking the Open-Source Type Ia
  Supernova Model to Longer Wavelengths for Next-Generation Cosmological
  Measurements, \mn@doi{10.48550/ARXIV.2209.05594}, \url
  {https://arxiv.org/abs/2209.05594}

\bibitem[\protect\citeauthoryear{{Planck Collaboration} et~al.,}{{Planck
  Collaboration} et~al.}{2016}]{planck2015}
{Planck Collaboration} et~al., 2016, \mn@doi [\aap]
  {10.1051/0004-6361/201525830}, \href
  {https://ui.adsabs.harvard.edu/abs/2016A&A...594A..13P} {594, A13}

\bibitem[\protect\citeauthoryear{{Planck Collaboration} et~al.,}{{Planck
  Collaboration} et~al.}{2020}]{planck_2018}
{Planck Collaboration} et~al., 2020, \mn@doi [\aap]
  {10.1051/0004-6361/201833910}, \href
  {https://ui.adsabs.harvard.edu/abs/2020A&A...641A...6P} {641, A6}

\bibitem[\protect\citeauthoryear{Popovic, Brout, Kessler, Scolnic  \&
  Lu}{Popovic et~al.}{2021}]{Popovic_2021}
Popovic B.,  Brout D.,  Kessler R.,  Scolnic D.,   Lu L.,  2021, \mn@doi [The
  Astrophysical Journal] {10.3847/1538-4357/abf14f}, 913, 49

\bibitem[\protect\citeauthoryear{{Pruzhinskaya}, {Novinskaya}, {Pauna}  \&
  {Rosnet}}{{Pruzhinskaya} et~al.}{2020}]{Pruzhinskaya_20}
{Pruzhinskaya} M.~V.,  {Novinskaya} A.~K.,  {Pauna} N.,   {Rosnet} P.,  2020,
  \mn@doi [\mnras] {10.1093/mnras/staa3173}, \href
  {https://ui.adsabs.harvard.edu/abs/2020MNRAS.499.5121P} {499, 5121}

\bibitem[\protect\citeauthoryear{{Riess} et~al.,}{{Riess}
  et~al.}{2016}]{riesshubb}
{Riess} A.~G.,  et~al., 2016, \mn@doi [\apj] {10.3847/0004-637X/826/1/56},
  \href {http://adsabs.harvard.edu/abs/2016ApJ...826...56R} {826, 56}

\bibitem[\protect\citeauthoryear{Riess et~al.,}{Riess
  et~al.}{2018}]{Riess_2018_highzSN}
Riess A.~G.,  et~al., 2018, \mn@doi [The Astrophysical Journal]
  {10.3847/1538-4357/aaa5a9}, 853, 126

\bibitem[\protect\citeauthoryear{{Riess} et~al.,}{{Riess}
  et~al.}{2022}]{riess_22}
{Riess} A.~G.,  et~al., 2022, \mn@doi [\apjl] {10.3847/2041-8213/ac5c5b}, \href
  {https://ui.adsabs.harvard.edu/abs/2022ApJ...934L...7R} {934, L7}

\bibitem[\protect\citeauthoryear{{Rigault} et~al.,}{{Rigault}
  et~al.}{2013}]{rigault_2013}
{Rigault} M.,  et~al., 2013, \mn@doi [\aap] {10.1051/0004-6361/201322104},
  \href {https://ui.adsabs.harvard.edu/abs/2013A&A...560A..66R} {560, A66}

\bibitem[\protect\citeauthoryear{{Rigault} et~al.,}{{Rigault}
  et~al.}{2015}]{393}
{Rigault} M.,  et~al., 2015, \mn@doi [\apj] {10.1088/0004-637X/802/1/20}, \href
  {http://adsabs.harvard.edu/abs/2015ApJ...802...20R} {802, 20}

\bibitem[\protect\citeauthoryear{{Rigault} et~al.,}{{Rigault}
  et~al.}{2020}]{rigault_20}
{Rigault} M.,  et~al., 2020, \mn@doi [\aap] {10.1051/0004-6361/201730404},
  \href {https://ui.adsabs.harvard.edu/abs/2020A&A...644A.176R} {644, A176}

\bibitem[\protect\citeauthoryear{Rodney et~al.,}{Rodney
  et~al.}{2014}]{Rodney_2014_highzSN}
Rodney S.~A.,  et~al., 2014, \mn@doi [The Astronomical Journal]
  {10.1088/0004-6256/148/1/13}, 148, 13

\bibitem[\protect\citeauthoryear{{Roman} et~al.,}{{Roman}
  et~al.}{2018}]{roman_18}
{Roman} M.,  et~al., 2018, \mn@doi [\aap] {10.1051/0004-6361/201731425}, \href
  {https://ui.adsabs.harvard.edu/abs/2018A&A...615A..68R} {615, A68}

\bibitem[\protect\citeauthoryear{{Rose}, {Garnavich}  \& {Berg}}{{Rose}
  et~al.}{2019}]{Rose_2019}
{Rose} B.~M.,  {Garnavich} P.~M.,   {Berg} M.~A.,  2019, \mn@doi [\apj]
  {10.3847/1538-4357/ab0704}, \href
  {https://ui.adsabs.harvard.edu/abs/2019ApJ...874...32R} {874, 32}

\bibitem[\protect\citeauthoryear{{Rose}, {Rubin}, {Strolger}  \&
  {Garnavich}}{{Rose} et~al.}{2021}]{rose_21_mass}
{Rose} B.~M.,  {Rubin} D.,  {Strolger} L.,   {Garnavich} P.~M.,  2021, \mn@doi
  [\apj] {10.3847/1538-4357/abd550}, \href
  {https://ui.adsabs.harvard.edu/abs/2021ApJ...909...28R} {909, 28}

\bibitem[\protect\citeauthoryear{{Saunders} et~al.,}{{Saunders}
  et~al.}{2018}]{snemo}
{Saunders} C.,  et~al., 2018, \mn@doi [\apj] {10.3847/1538-4357/aaec7e}, \href
  {https://ui.adsabs.harvard.edu/abs/2018ApJ...869..167S} {869, 167}

\bibitem[\protect\citeauthoryear{{Scolnic} et~al.,}{{Scolnic}
  et~al.}{2018}]{panstarrs18}
{Scolnic} D.~M.,  et~al., 2018, \mn@doi [\apj] {10.3847/1538-4357/aab9bb},
  \href {http://adsabs.harvard.edu/abs/2018ApJ...859..101S} {859, 101}

\bibitem[\protect\citeauthoryear{{Scolnic} et~al.,}{{Scolnic}
  et~al.}{2022}]{panplus_datarelease}
{Scolnic} D.,  et~al., 2022, \mn@doi [\apj] {10.3847/1538-4357/ac8b7a}, \href
  {https://ui.adsabs.harvard.edu/abs/2022ApJ...938..113S} {938, 113}

\bibitem[\protect\citeauthoryear{{Siebert}, {Foley}, {Jones}  \&
  {Davis}}{{Siebert} et~al.}{2020}]{siebert_2020}
{Siebert} M.~R.,  {Foley} R.~J.,  {Jones} D.~O.,   {Davis} K.~W.,  2020,
  \mn@doi [\mnras] {10.1093/mnras/staa577}, \href
  {https://ui.adsabs.harvard.edu/abs/2020MNRAS.493.5713S} {493, 5713}

\bibitem[\protect\citeauthoryear{Smith et~al.,}{Smith
  et~al.}{2020}]{smith2020cosmology}
Smith M.,  et~al., 2020, First Cosmology Results using Type Ia Supernovae from
  the Dark Energy Survey: The Effect of Host Galaxy Properties on Supernova
  Luminosity (\mn@eprint {arXiv} {2001.11294})

\bibitem[\protect\citeauthoryear{{Steigerwald}, {Rodrigues}, {Profumo}  \&
  {Marra}}{{Steigerwald} et~al.}{2022}]{steigerwald_22}
{Steigerwald} H.,  {Rodrigues} D.,  {Profumo} S.,   {Marra} V.,  2022, \mn@doi
  [\mnras] {10.1093/mnras/stab3747}, \href
  {https://ui.adsabs.harvard.edu/abs/2022MNRAS.510.4779S} {510, 4779}

\bibitem[\protect\citeauthoryear{{Sullivan} et~al.,}{{Sullivan}
  et~al.}{2010}]{sullivan_host}
{Sullivan} M.,  et~al., 2010, \mn@doi [\mnras]
  {10.1111/j.1365-2966.2010.16731.x}, \href
  {https://ui.adsabs.harvard.edu/abs/2010MNRAS.406..782S} {406, 782}

\bibitem[\protect\citeauthoryear{{Suzuki} et~al.,}{{Suzuki}
  et~al.}{2012}]{suzuki12}
{Suzuki} N.,  et~al., 2012, \mn@doi [\apj] {10.1088/0004-637X/746/1/85}, \href
  {https://ui.adsabs.harvard.edu/abs/2012ApJ...746...85S} {746, 85}

\bibitem[\protect\citeauthoryear{Taylor, Lidman, Tucker, Brout, Hinton  \&
  Kessler}{Taylor et~al.}{2021}]{T21}
Taylor G.,  Lidman C.,  Tucker B.~E.,  Brout D.,  Hinton S.~R.,   Kessler R.,
  2021, \mn@doi [Monthly Notices of the Royal Astronomical Society]
  {10.1093/mnras/stab962}, 504, 4111

\bibitem[\protect\citeauthoryear{{Taylor} et~al.,}{{Taylor} et~al.}{2023}]{t23}
{Taylor} G.,  et~al., 2023, \mn@doi [\mnras] {10.1093/mnras/stad320}, \href
  {https://ui.adsabs.harvard.edu/abs/2023MNRAS.tmp..345T} {}

\bibitem[\protect\citeauthoryear{{Tripp}}{{Tripp}}{1998}]{tripp98}
{Tripp} R.,  1998, \aap, \href
  {https://ui.adsabs.harvard.edu/abs/1998A&A...331..815T} {331, 815}

\bibitem[\protect\citeauthoryear{{Wiseman} et~al.,}{{Wiseman}
  et~al.}{2020}]{wiseman_2020}
{Wiseman} P.,  et~al., 2020, \mn@doi [\mnras] {10.1093/mnras/staa1302}, \href
  {https://ui.adsabs.harvard.edu/abs/2020MNRAS.495.4040W} {495, 4040}

\bibitem[\protect\citeauthoryear{{Wiseman} et~al.,}{{Wiseman}
  et~al.}{2022}]{wiseman_22}
{Wiseman} P.,  et~al., 2022, \mn@doi [\mnras] {10.1093/mnras/stac1984}, \href
  {https://ui.adsabs.harvard.edu/abs/2022MNRAS.515.4587W} {515, 4587}

\makeatother
\end{thebibliography}

%%%%%%%%%%%%%%%%%%%%%%%%%%%%%%%%%%%%%%%%%%%%%%%%%%

%%%%%%%%%%%%%%%%% APPENDICES %%%%%%%%%%%%%%%%%%%%%

\appendix

\section{The Cosmological Impact of UV Ringing in SALT3.LOWMASS}
\label{4000_analysis}

We repeat our cosmological analysis over an abridged rest-frame wavelength range of $4000 < \lambda_{\rm eff} < 8000$\AA, in order to subtract out any effects of non-physical ringing in the UV region of our SALT3.LOWMASS surface. The results reported here will also include the unavoidable effects of fitting light curve models and distances on fewer data points, although that effect should impact all surfaces to the same degree.

\begin{figure*}
  \begin{center}
   \includegraphics[width=0.95\columnwidth]{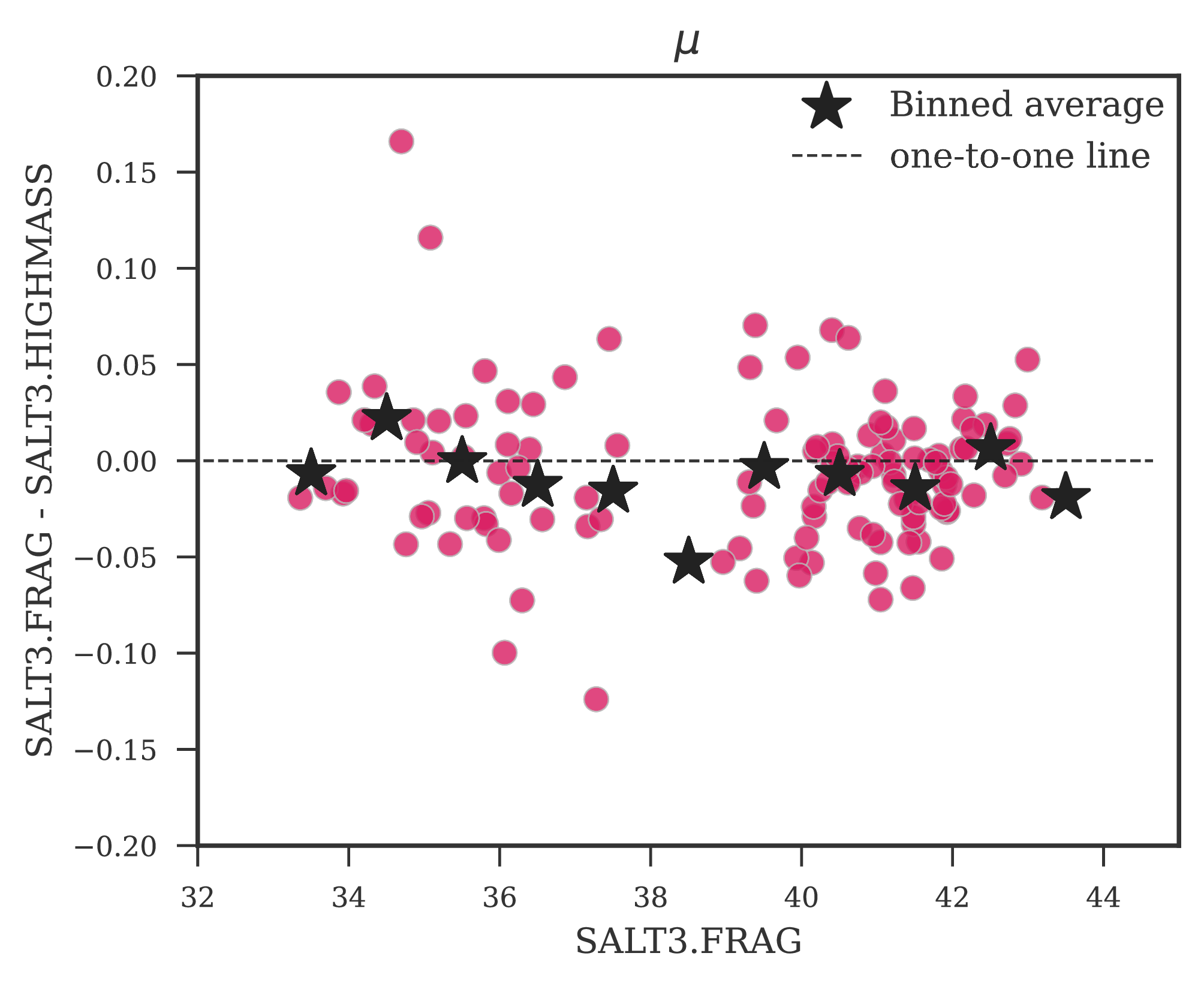}
   \includegraphics[width=0.95\columnwidth]{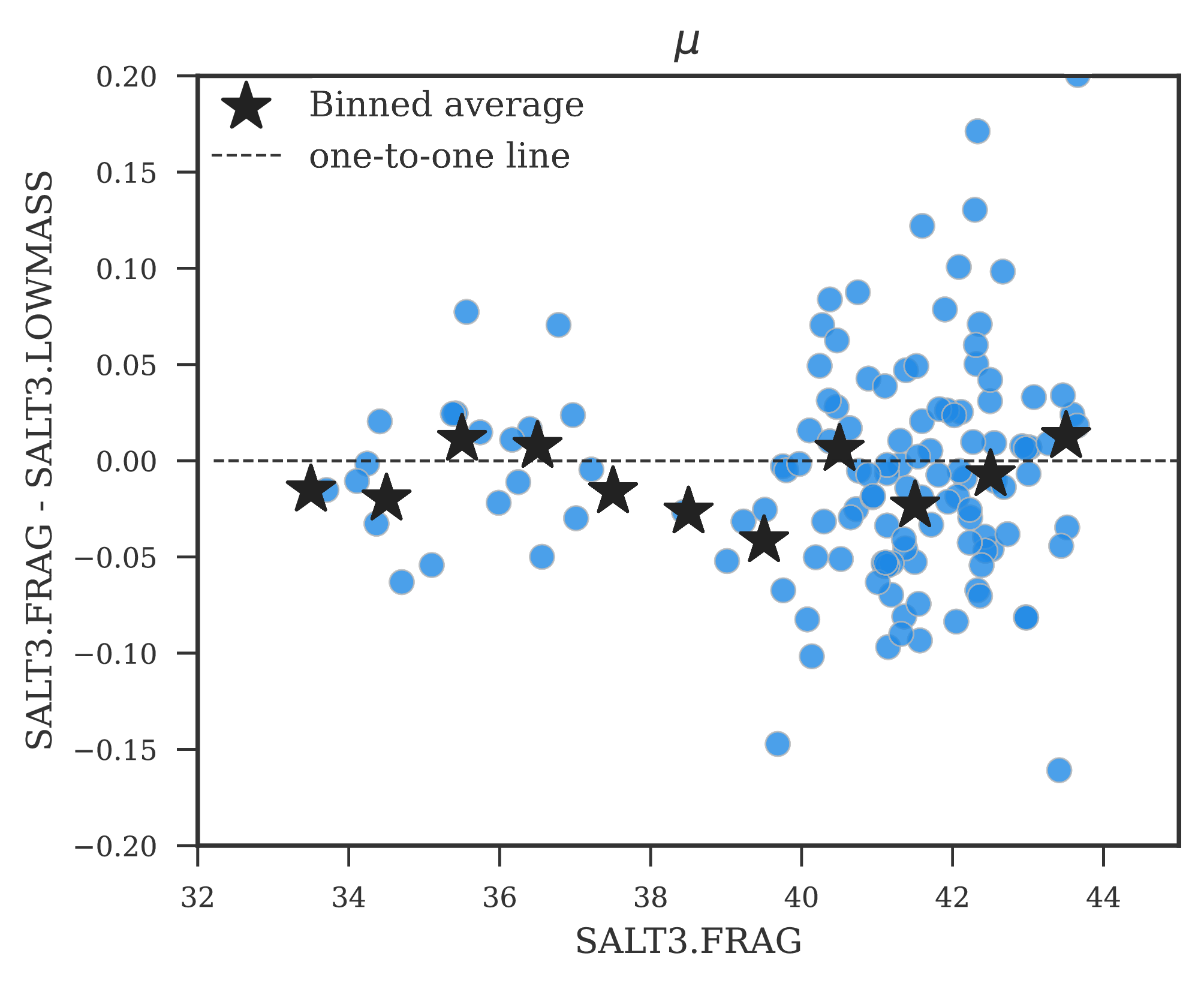}
  \end{center}
  \caption{Comparison of the distance moduli for the DES-SN3YR sample when fit over an abridged range of $4000\leq \lambda_{\rm rest} \leq 8000$\AA\ with SALT3.FRAG (including a mass step) versus SALT3.HIGHMASS (\textit{left panel, pink points}) and SALT3.LOWMASS (\textit{right panel, blue points}). We plot the difference in distance moduli on the y-axis.}
  \label{mu_noUV}
\end{figure*}

In Figure~\ref{mu_noUV} we plot the change in fitted $\mu$ between the SALT3.FRAG and SALT3.LOWMASS/HIGHMASS surfaces. We see that the slight distance-dependent slope in the recovery of $\mu$ between SALT3.LOWMASS and SALT3.FRAG in our nominal analysis (Figure~\ref{lcfit_mu}) is no longer present when using the reduced wavelength range. We do not observe any such change in slope in the recovery of $\mu$ between SALT3.HIGHMASS and SALT3.FRAG.

In Table~\ref{mass_sample_cosmo_noUV}, we report the best fit parameters from Equation~\ref{eq:modified_tripp_p3}. There are no statistically significant changes in the recovered parameters between this analysis and our nominal analysis (Table~\ref{mass_sample_cosmo}), though we do lose four low-host-mass SNe and five high-host-mass SNe from our samples due to fitting cuts.

In Table~\ref{mass_sample_HR_noUV}, we report the scatter in the Hubble residual from the best-fit cosmology for each surface. All surfaces see an increase in Hubble residual scatter compared to the nominal analysis (Table~\ref{cosmo_sigma}). However, the \textit{relationships} between the Hubble residual scatter across different surfaces has not changed significantly from the nominal analysis. While in some cases the Hubble residual scatter now appears to be worse with SALT3.LOWMASS than SALT3.FRAG when considering only SNe from low-mass hosts, the statistical significance of this is low; 
the Hubble residual scatter decreases by $0.6-0.7\sigma$ when using SALT3.LOWMASS compared to SALT3.FRAG in the nominal analysis, and similarly changes by $\pm 0.1\sigma$ in the analysis without UV data.

\begin{table*}
\begin{minipage}{\textwidth}
  \renewcommand{\arraystretch}{1.5}
  \centering
  \caption{The best fit nuisance parameters (from Equation~\ref{eq:modified_tripp_p3}) and cosmology parameters for the DES-SN3YR sample when fit with different SALT3 surfaces, when using a reduced wavelength range for light curve fitting.}
    \label{mass_sample_cosmo_noUV}
\begin{tabular}{lllllllllll}
\hline
Sample & \# SNe fit & $\alpha$ & $\beta$ & $\sigma_{\rm int}$ & $M$ & $\Omega_M$ & $\sigma_{\Omega_M}^{\rm stat}$ & $w$ & $\sigma_w^{\rm stat}$ & $\chi^2_{\rm cosmo}$ \\ \hline
SALT3.FRAG & 314 & 0.14 $\pm$ 0.01 & 2.74 $\pm$ 0.12 & 0.12 & -29.98 & 0.33 & 0.02 & -0.96 & 0.05 & 8.3 \\
SALT3.FRAG + Mass Step & 314 & 0.16 $\pm$ 0.01 & 2.76 $\pm$ 0.11 & 0.11 & -29.97 & 0.35 & 0.02 & -0.89 & 0.05 & 7.8 \\
SALT3.HIGHMASS & 124 & 0.13 $\pm$ 0.01 & 2.63 $\pm$ 0.17 & 0.09 & -29.93 & 0.32 & 0.02 & -0.99 & 0.07 & 9.3 \\
SALT3.LOWMASS & 130 & 0.08 $\pm$ 0.02 & 2.84 $\pm$ 0.18 & 0.11 & -30.09 & 0.35 & 0.02 & -0.88 & 0.09 & 11.8 \\\hline
\end{tabular}
\end{minipage}
\end{table*}

\begin{table*}
\begin{minipage}{\textwidth}
  \renewcommand{\arraystretch}{1.5}
  \centering
  \caption{Scatter in the Hubble residual (from best-fit cosmology) for each SALT3 surface, when using a reduced wavelength range for light curve fitting.. All values are in units of magnitude. The best-fit cosmology is unique for each entry in the ``Result" column}
    \label{mass_sample_HR_noUV}
\begin{tabular}{llll}
\hline
\multicolumn{4}{c}{\textit{Hubble residuals from best-fit cosmology}} \\ \hline
Result & Subsample & RMS$_{\rm HR}$ & Weighted RMS$_{\rm HR}$ \\ \hline
 & Full & 0.169$\pm$0.008 & 0.158$\pm$0.008 \\
 & \cellcolor[HTML]{EFA1BE}$M_{\log} > 10$ & \cellcolor[HTML]{EFA1BE}0.154$\pm$0.009 & \cellcolor[HTML]{EFA1BE}0.145$\pm$0.010 \\
\multirow{-3}{*}{SALT3.FRAG} & \cellcolor[HTML]{90C7F7}$M_{\log} < 10$ & \cellcolor[HTML]{90C7F7}0.188$\pm$0.012 & \cellcolor[HTML]{90C7F7}0.174$\pm$0.014 \\ \hline
 & Full & 0.171$\pm$0.008 & 0.158$\pm$0.008 \\
 & \cellcolor[HTML]{EFA1BE}$M_{\log} > 10$ & \cellcolor[HTML]{EFA1BE}0.157$\pm$0.010 & \cellcolor[HTML]{EFA1BE}0.147$\pm$0.010 \\
\multirow{-3}{*}{\begin{tabular}[c]{@{}l@{}}SALT3.FRAG\\  + Mass Step\end{tabular}} & \cellcolor[HTML]{90C7F7}$M_{\log} < 10$ & \cellcolor[HTML]{90C7F7}0.188$\pm$0.012 & \cellcolor[HTML]{90C7F7}0.173$\pm$0.014 \\ \hline
SALT3.HIGHMASS & \cellcolor[HTML]{EFA1BE}$M_{\log} > 10$ & \cellcolor[HTML]{EFA1BE}0.146$\pm$0.012 & \cellcolor[HTML]{EFA1BE}0.132$\pm$0.013 \\ \hline
SALT3.LOWMASS & \cellcolor[HTML]{90C7F7}$M_{\log} < 10$ & \cellcolor[HTML]{90C7F7}0.19$\pm$0.013 & \cellcolor[HTML]{90C7F7}0.172$\pm$0.014 \\ \hline
\end{tabular}
\end{minipage}
\end{table*}

% Don't change these lines
\bsp	% typesetting comment
\label{lastpage}
\end{document}